\documentclass[preprint2,longabstract]{aastex}

\usepackage[utf8]{inputenc}
\usepackage[english]{babel}
\usepackage{graphicx}
\usepackage{lineno}
\usepackage{natbib}
\usepackage[mathcal]{eucal}
\usepackage{mathrsfs}

\newcommand{\nhi}{N(\mathrm{H\,\scriptstyle{I}})}

\newcommand{\nhd}{N(\mathrm{H_2})}
\newcommand{\wco}{W_\mathrm{CO}}
\newcommand{\hi}{\mathrm{H\,\scriptstyle{I}}}
\newcommand{\hii}{\mathrm{H\,\scriptstyle{II}}}
\newcommand{\hd}{\mathrm{H}_2}
\newcommand{\xco}{X_\mathrm{CO}}
\newcommand{\ebv}{\mathrm{E(B-V)}}
\newcommand{\ebvres}{\ebv_\mathrm{res}}
\newcommand{\xebv}{X_\mathrm{EBV}}
\newcommand{\nef}{\epsilon_\mathrm{N}}
\newcommand{\qhi}[1]{q_{\mathrm{H\,\scriptscriptstyle{I}} , \, #1}}
\newcommand{\qhinull}{q_\mathrm{H\,\scriptscriptstyle{I}}}
\newcommand{\qco}[1]{q_{\mathrm{CO} , \, #1}}
\newcommand{\qconull}{q_\mathrm{CO}}
\newcommand{\qebv}{q_\mathrm{EBV}}
\newcommand{\like}{\mathscr{L}}

\shorttitle{{Fermi} observations of Cas & Cep}
\shortauthors{Abdo et al.}

\begin{document}


\title{Fermi observations of Cassiopeia and Cepheus: diffuse gamma-ray emission
in the outer Galaxy
}
\author{
A.~A.~Abdo\altaffilmark{2,3}, 
M.~Ackermann\altaffilmark{4}, 
M.~Ajello\altaffilmark{4}, 
L.~Baldini\altaffilmark{5}, 
J.~Ballet\altaffilmark{6}, 
G.~Barbiellini\altaffilmark{7,8}, 
D.~Bastieri\altaffilmark{9,10}, 
B.~M.~Baughman\altaffilmark{11}, 
K.~Bechtol\altaffilmark{4}, 
R.~Bellazzini\altaffilmark{5}
% ...
}
% ...
\altaffiltext{2}{Space Science Division, Naval Research Laboratory, Washington,
DC 20375, USA}
\altaffiltext{3}{National Research Council Research Associate, National Academy
of Sciences, Washington, DC 20001, USA}
\altaffiltext{4}{W. W. Hansen Experimental Physics Laboratory, Kavli Institute
for Particle Astrophysics and Cosmology, Department of Physics and SLAC National
Accelerator Laboratory, Stanford University, Stanford, CA 94305, USA}
\altaffiltext{5}{Istituto Nazionale di Fisica Nucleare, Sezione di Pisa, I-56127
Pisa, Italy}
\altaffiltext{6}{Laboratoire AIM, CEA-IRFU/CNRS/Universit\'e Paris Diderot,
Service d'Astrophysique, CEA Saclay, 91191 Gif sur Yvette, France}
\altaffiltext{7}{Istituto Nazionale di Fisica Nucleare, Sezione di Trieste,
I-34127 Trieste, Italy}
\altaffiltext{8}{Dipartimento di Fisica, Universit\`a di Trieste, I-34127
Trieste, Italy}
\altaffiltext{9}{Istituto Nazionale di Fisica Nucleare, Sezione di Padova,
I-35131 Padova, Italy}
\altaffiltext{10}{Dipartimento di Fisica ``G. Galilei", Universit\`a di Padova,
I-35131 Padova, Italy}
\altaffiltext{11}{Department of Physics, Center for Cosmology and Astro-Particle
Physics, The Ohio State University, Columbus, OH 43210, USA}
% ...

\begin{abstract}
We present the analysis of the interstellar $\gamma$-ray emission measured by the \emph{Fermi} Large Area Telescope toward a region in the second Galactic quadrant at $100^\circ \leq l \leq 145^\circ$ and $-15^\circ \leq b \leq +30^\circ$. This region encompasses the prominent Gould-Belt clouds of Cassiopeia, Cepheus and the Polaris flare, as well as atomic and molecular complexes at larger distances, like that associated with NGC 7538 in the Perseus arm. The good kinematic separation in velocity between the local, Perseus, and outer arms, and the presence of massive complexes in each of them make this region well suited to probe cosmic rays and the interstellar medium beyond the solar circle. The $\gamma$-ray emissivity spectrum of the gas in the Gould Belt is consistent with expectations based on the locally measured cosmic-ray spectra. The $\gamma$-ray emissivity decreases from the Gould Belt to the Perseus arm, but the measured gradient is flatter than expectations for cosmic-ray sources peaking in the inner Galaxy as suggested by pulsars. The $\xco=\nhd/\wco$ conversion factor is found to increase from $(0.87 \pm 0.05) \times 10^{20}$ cm$^{-2}$ (K km s$^{-1}$)$^{-1}$ in the Gould Belt to $(1.9\pm 0.2) \times 10^{20}$ cm$^{-2}$ (K km s$^{-1}$)$^{-1}$ in the Perseus arm. We derive masses for the molecular clouds under study. Dark gas, not properly traced by radio and microwave surveys, is detected in the Gould Belt through a correlated excess of dust and $\gamma$-ray emission: its mass amounts to $\sim 50\%$ of the CO-traced mass.
\end{abstract}


\section*{}
\clearpage

\section{Introduction}\label{intro}
Galactic interstellar $\gamma$-ray emission is produced through the interactions
of high-energy cosmic rays (CRs) with the gas in the interstellar medium (ISM)
(via pion production and Bremsstrahlung) and with the interstellar radiation
field (via Inverse Compton, IC, scattering). Thus, since early studies with the
COS-B satellite, diffuse $\gamma$ rays were recognized to be a tracer of the CR
densities and of ISM column densities in the Galaxy \citep{lebrun, gradient,
cosbdiff}.

The interpretation of the observed emission is often based on two radio tracers of the interstellar gas: the 21 cm line of the hyperfine transition of atomic hydrogen ($\hi$) is used to derive its column density $\nhi$; the 2.6 mm line of the rotational transition $\mathrm{J}=1\rightarrow 0$ of CO is used to trace the molecular gas.  The molecular phase of the ISM is composed mainly of $\hd$ which cannot be traced directly in its cold phase. It has long been verified, primarily using  virial mass estimates, that the brightness temperature of CO integrated over velocity, $\wco$, roughly scales with the total molecular mass in the emitting region \citep[see e.g.][]{solomon}. The conversion factor that transforms $\wco$ into $\hd$ column density is known as $\xco=\nhd / \wco$ \citep{lebrun}.

The $\xco$ conversion factor has often been assumed to be uniform across the Galaxy. We now have evidence, however, that it should increase in the outer Galaxy: from virial masses \citep{sethvirial}, from COBE/DIRBE studies \citep{cobedirbe1,cobedirbe2} and from the measurement of the Galactic metallicity gradient \citep{israel1,israel2}. A precise estimate of the $\xco$ gradient is necessary to measure the masses of distant $\hd$ clouds, but it also impacts the derivation of the distribution of cosmic-ray sources from $\gamma$-ray observations \citep{strongXvar}.

For many years supernova remnants (SNRs) have been considered the best candidates as CR sources. We have recently detected possible signatures of hadron acceleration in SNRs thanks to $\gamma$-ray observations in the TeV \citep{magicSNR,hessSNR,veritasSNR} and GeV domain \citep{LATsnr}. However, the origin of Galactic cosmic rays is still mysterious and, on the other hand, the distribution of SNRs in the Galaxy is very poorly determined \citep{snrunc}, leading to large uncertainties in the models of diffuse $\gamma$-ray emission. The $\gamma$-ray emissivity gradient of the diffuse $\hi$ gas can provide useful constraints on the CR density distribution.

Since the Doppler shift of the radio lines allows kinematic separation of different structures along a line of sight, it is possible to constrain the $\gamma$-ray emissivities and the subsequent $\xco$ ratios in specific Galactic regions. The performance of the previous $\gamma$-ray telescopes did not allow very precise measurements beyond the solar circle \citep{egretcep,egretmono}. The situation has recently been improved with the successful launch of the \emph{Fermi} Gamma-ray Space Telescope on 2008 June 11. The Large Area Telescope (LAT) on board the \emph{Fermi} mission \citep{latpaper} has a sensitivity more than an order of magnitude greater than the previous instrument EGRET on board the Compton Gamma-Ray Observatory and a superior angular resolution.

We present here the analysis of the interstellar $\gamma$ radiation measured by the \emph{Fermi} LAT in a selected region of the second Galactic quadrant, at $100^\circ \leq l \leq 145^\circ$, $-15^\circ \leq b \leq +30^\circ$, during the first 11 months of the science phase of the mission. The region was chosen because here the velocity gradient with Galactocentric distance is very steep, resulting in good kinematic separation which allows four different regions to be defined along each line of sight: the nearby Gould Belt, the main part of the local arm, and the more distant Perseus and outer spiral arms. Among the most conspicuous clouds, one finds Cassiopeia, the Cepheus and Polaris flares in the Gould Belt \citep{gouldbelt,polaris,greniercepcas}, the most massive molecular complex in the Perseus arm associated with NGC 7538 and Cas A \citep{ungerechts}, and the off-plane molecular cloud in the Perseus arm associated with NGC 281 \citep{sato}. These prominent cloud complexes are well suited to probe CRs and the ISM. The motivations of this work are both to provide improved constraints for diffuse emission models to be used in the detection and analysis of LAT sources and to reach a better comprehension of the physical phenomena related with diffuse $\gamma$-ray emission in the outer Galaxy.

\section{Interstellar gas}

Here we describe the preparation of the maps tracing the column densities of the different components of the ISM, used in the following section to analyse LAT data.

\subsection{Radio and Microwave data}

\subsubsection{$\hi$}\label{hidata}
Column densities $\nhi$ of atomic hydrogen have been derived using the LAB $\hi$ survey by \citet{labsurvey}. The LSR velocity\footnote{Local Standard of Rest velocity, i.e. the velocity in a reference frame following the motion of the solar system.} coverage spans from $- 450$ km~s$^{-1}$ to $+400$ km~s$^{-1}$ with a resolution of 1.3 km~s$^{-1}$. The survey angular resolution is about $0.6^\circ$. Owing to the strong absorption against the radio continuum emission of the Cas A supernova remnant, the $\hi$ column densities within $0.5^\circ$ from its position were determined by linear interpolation of the adjacent lines.

The column densities have been derived applying an optical depth correction for a uniform spin temperature $T_S = 125$ K, in order to directly compare our results with previous studies \citep[like][]{egretcep}. There is not general agreement in the literature about the values of the spin temperature in the atomic phase of the ISM. From observations of the 21 cm line of $\hi$ seen in absorption, \citet{mohanI,mohanII} derived for our region values of $T_S$ varying from $\sim 50$ K to $\gtrsim 2000$ K, with a mean value $\sim 125$ K. Recently \citet{dickey}, on the basis of other $\hi$ absorption surveys, reported a mean value in the second Galactic quadrant $T_S = 250$ K, almost constant with Galactocentric radius.  The maximum difference between the values of $\nhi$ obtained with $T_S = 125$ K and those obtained in the optically-thin approximation (corresponding to the lower possible amount of gas or to the limit of very high spin temperature) is $30\%-40\%$, whereas the maximum difference between $T_S = 250$ K and optically-thin approximation is $10\%-15\%$. The optical depth correction is non-linear, so assessing the effects of the approximation is not trivial: in particular we note that the uncertainties are larger where the gas density is higher and that assuming lower values for $T_S$ we obtain structured excesses in modeled diffuse $\gamma$-ray intensities following the shape of the clouds.

The systematic errors are even larger in the Galactic plane where self absorption phenomena become important, especially in the Perseus arm where the subsequent uncertainties of derived $\nhi$ can reach 30\% \citep{gibson}.

\subsubsection{CO}\label{codata}

Intensities $\wco$ of the 2.6 mm line of CO have been derived from the composite survey of \citet{cosurvey}, with sampling every 0.125$^\circ$ near the Galactic plane and in the Gould Belt clouds, supplemented with observations at 0.25$^\circ$ sampling for high-latitude clouds ($>5^\circ$), covering in particular the region of NGC 281.

Lines of sight not surveyed in CO were restored by linear interpolation of adjacent directions where possible; otherwise they were assumed to be free of significant CO emission. CO data have been filtered with the moment-masking technique in order to reduce the noise while keeping the resolution of the original data and retaining the edges of the CO clouds \citep[see e.g.][Section~2.3]{cosurvey}. Preserving the faint CO edges is important to help decrease the degree of spatial correlation that naturally exists between the $\nhi$ and $\wco$ maps of a given cloud complex because of the ISM multi-phase structure.

\subsection{Kinematic separation of the Galactic structures}\label{separation}
Our aim is to separately determine the $\gamma$-ray emission from the different Galactic structures present along the line of sight in the second quadrant:
\begin{enumerate}
\item the very nearby complexes in the Gould Belt, within $\sim 300$ pc from the solar system;
\item the main part of the local arm, typically $\sim 1$ kpc away;
\item the Perseus arm, 2.5 to 4 kpc away;
\item the outer arm and beyond.
\end{enumerate}

The separation between the Gould-Belt and local-arm components is important to probe for a possible change in cosmic-ray densities between the quiescent nearby clouds of Cassiopeia and Cepheus, that produce few low-mass stars, and the more active regions of the local arm which shelter several OB associations \citep{greniercepcas}.

\begin{figure}[!hbt]
\plotone{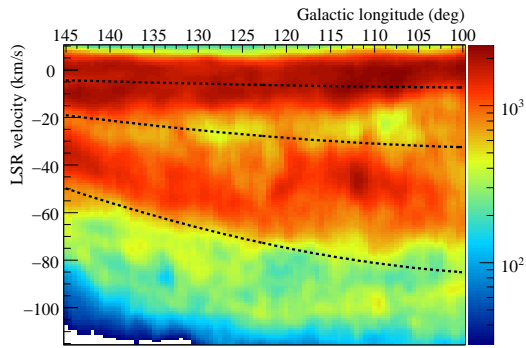}
\caption{$\hi$ longitude-velocity diagram obtained by integrating the brightness
temperature in \citet{labsurvey} for $|b|<10^\circ$. The color scale is
logarithmic in units of deg K. The three curves bound the preliminary
Galactocentric rings used for analysis. At $R=8.8$ kpc, $R=10$ kpc, and $R=14$
kpc (from top to bottom) they roughly separate Gould Belt, local arm, Perseus
and outer arm. The separation between Gould Belt and main part of the local arm
is hard to distinguish in this diagram.}\label{lvdiagram}
\end{figure}

The good kinematic separation of the interstellar gas in this part of the sky is
illustrated in Fig.~\ref{lvdiagram}.

\begin{figure*}[!p]
\begin{center}
\begin{tabular}{cc}
\includegraphics[width=0.47\textwidth]{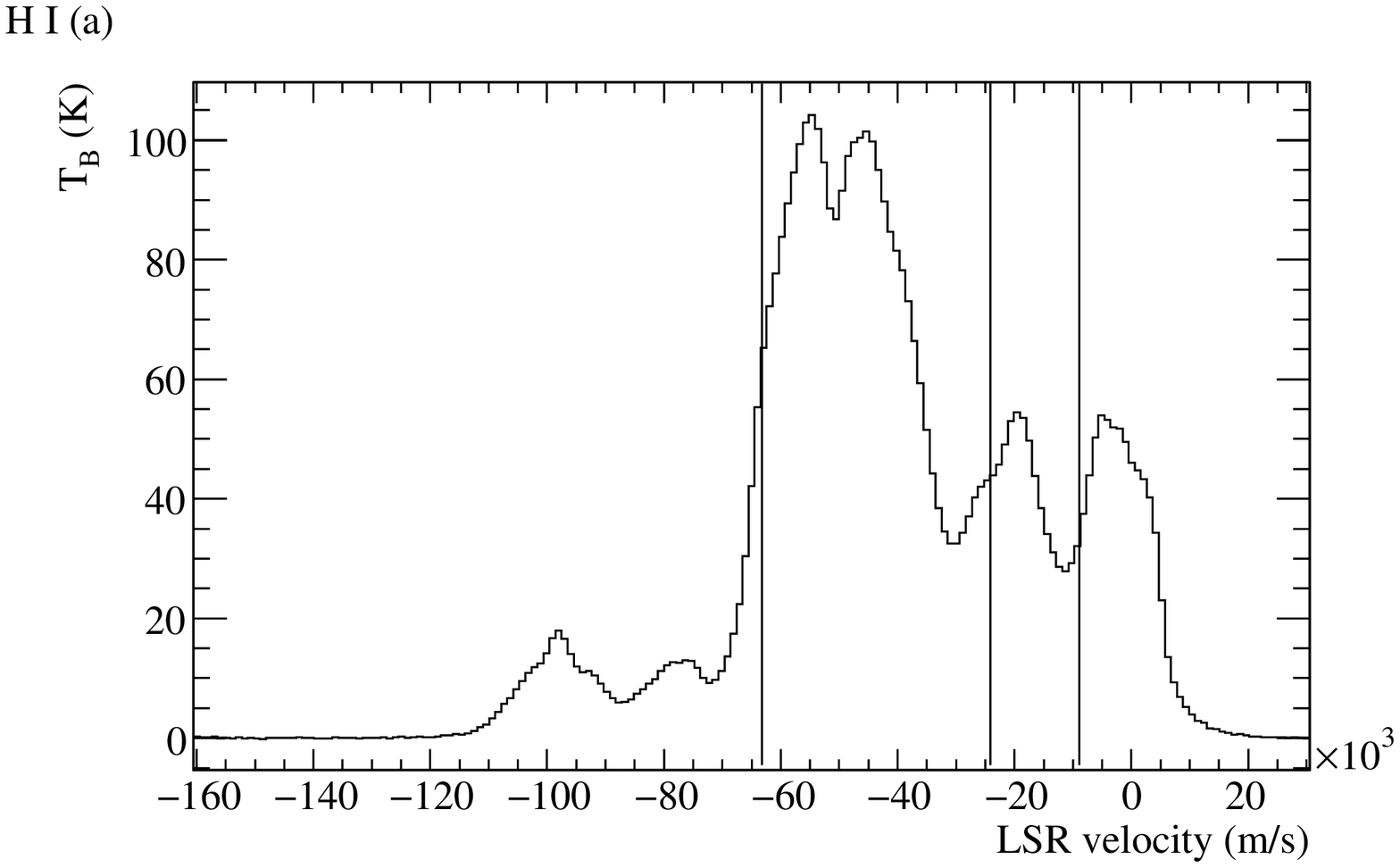} &
\includegraphics[width=0.47\textwidth]{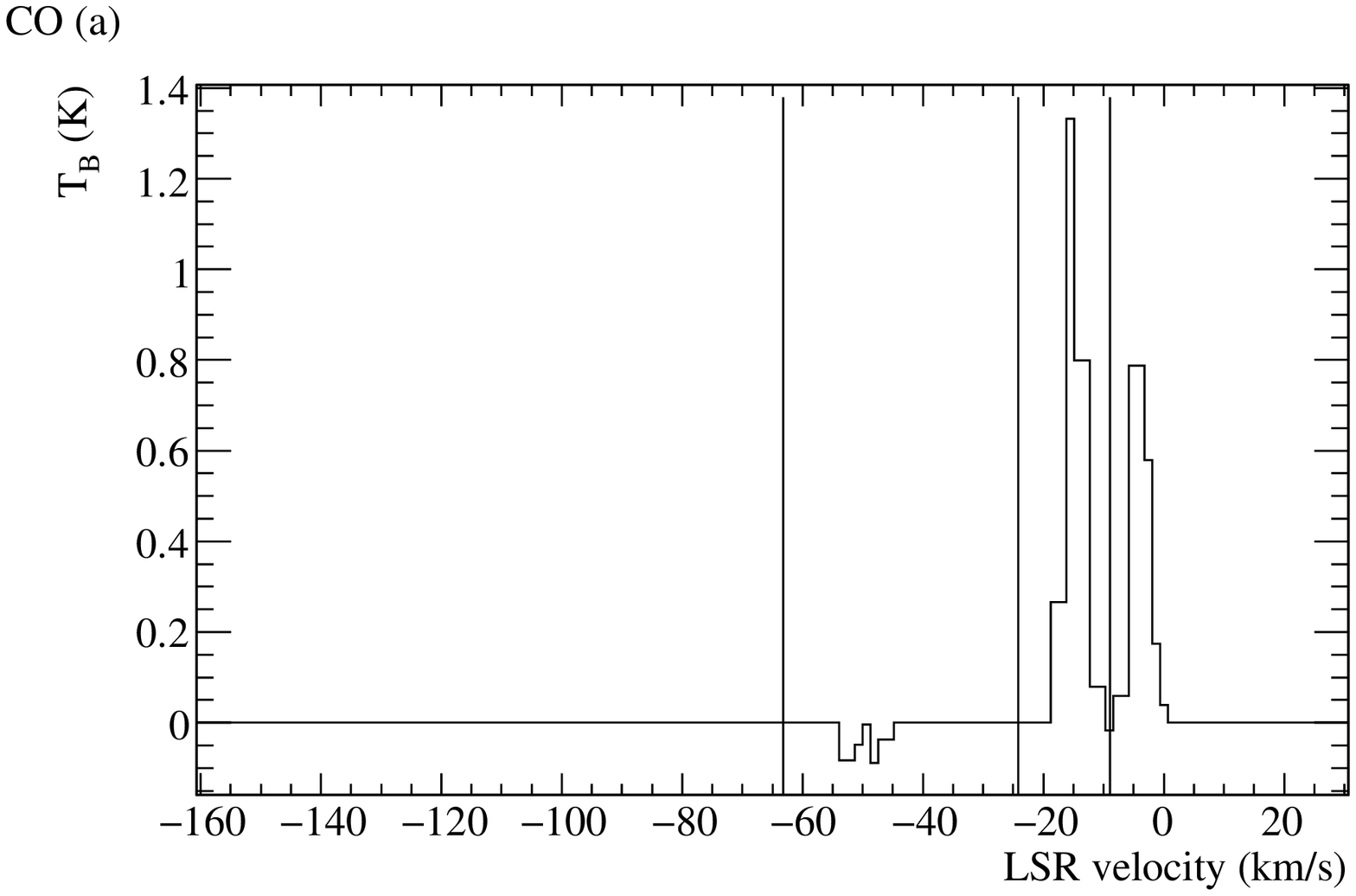}\\
\includegraphics[width=0.47\textwidth]{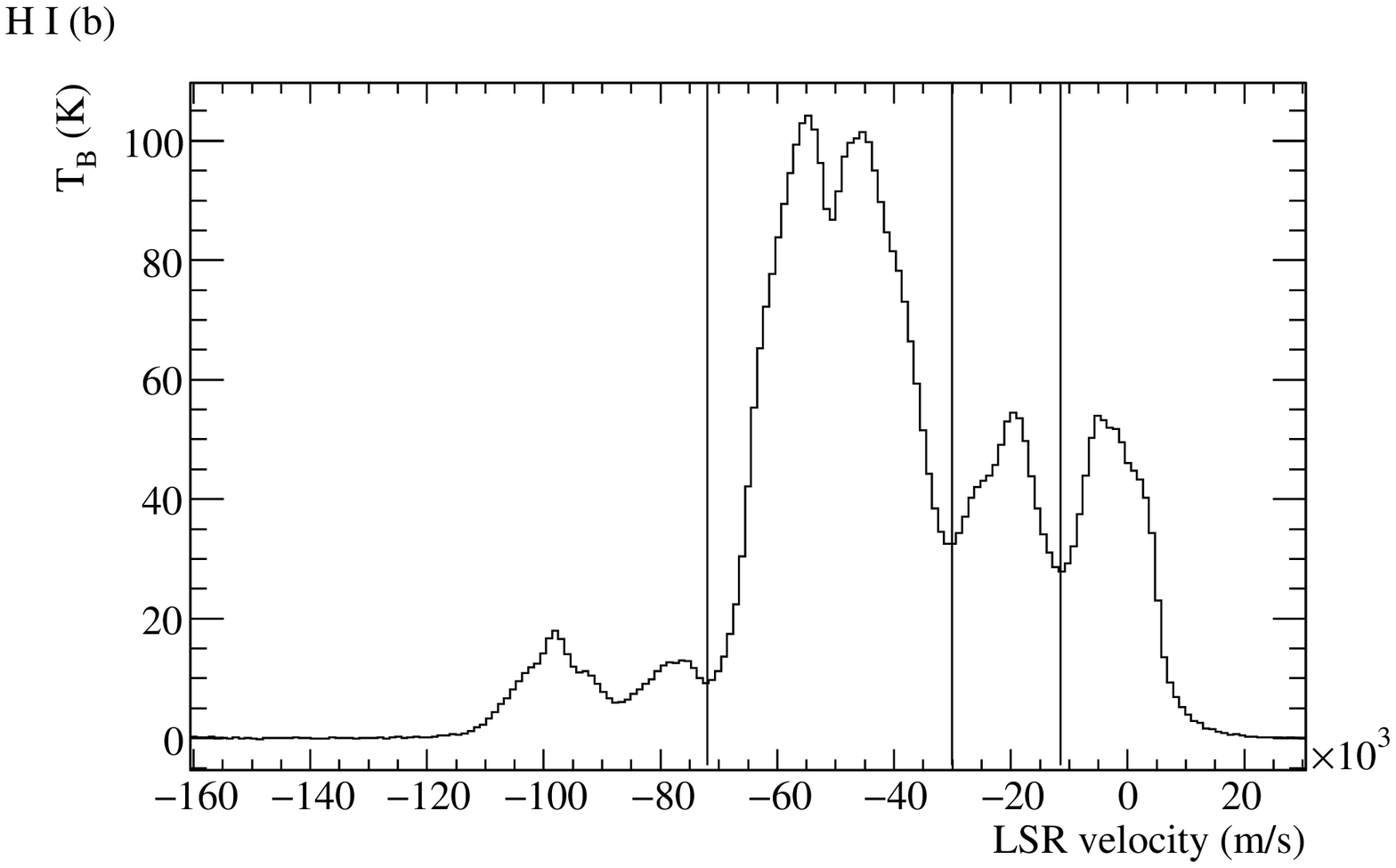} &
\includegraphics[width=0.47\textwidth]{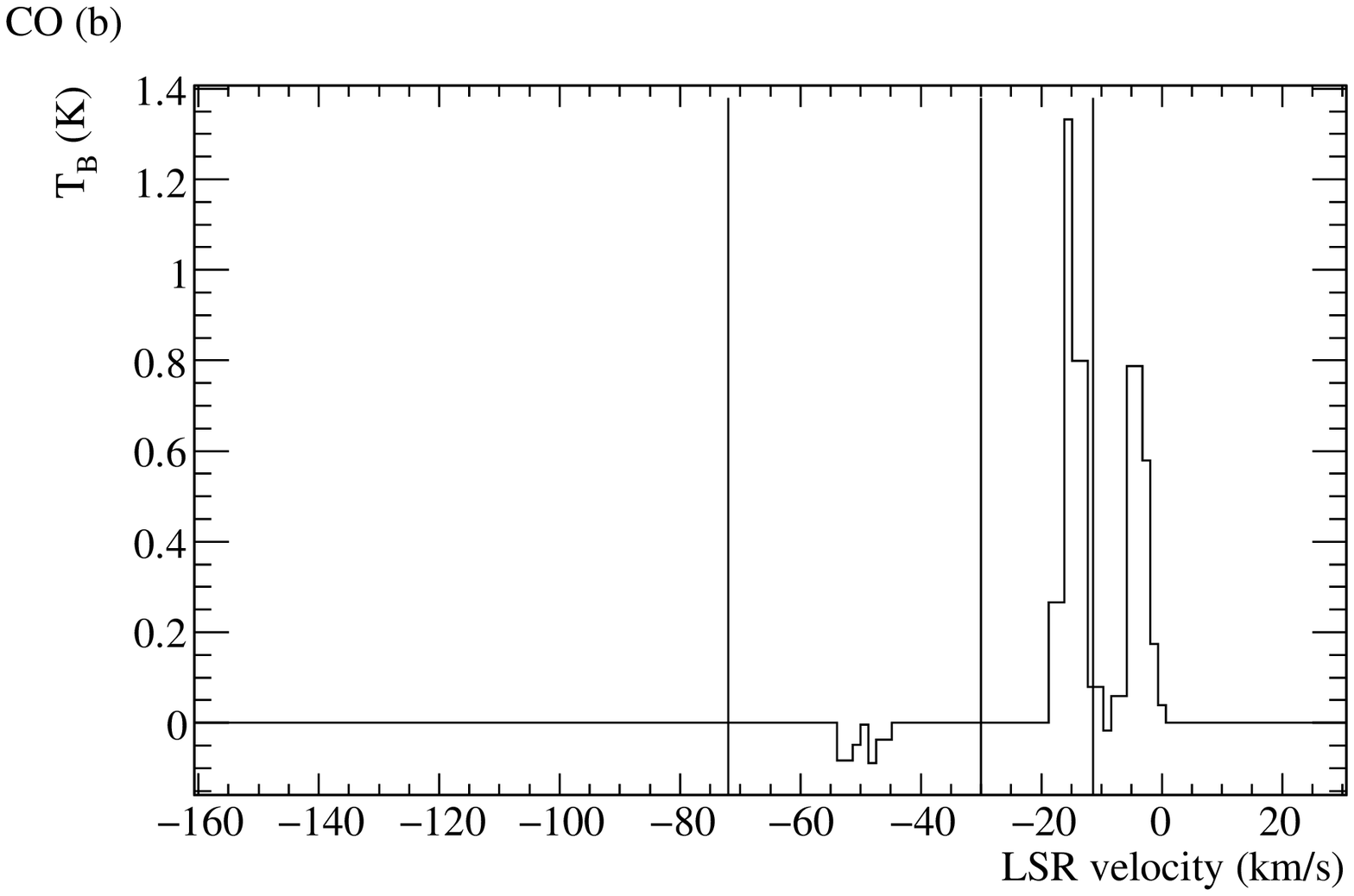} \\
\includegraphics[width=0.47\textwidth]{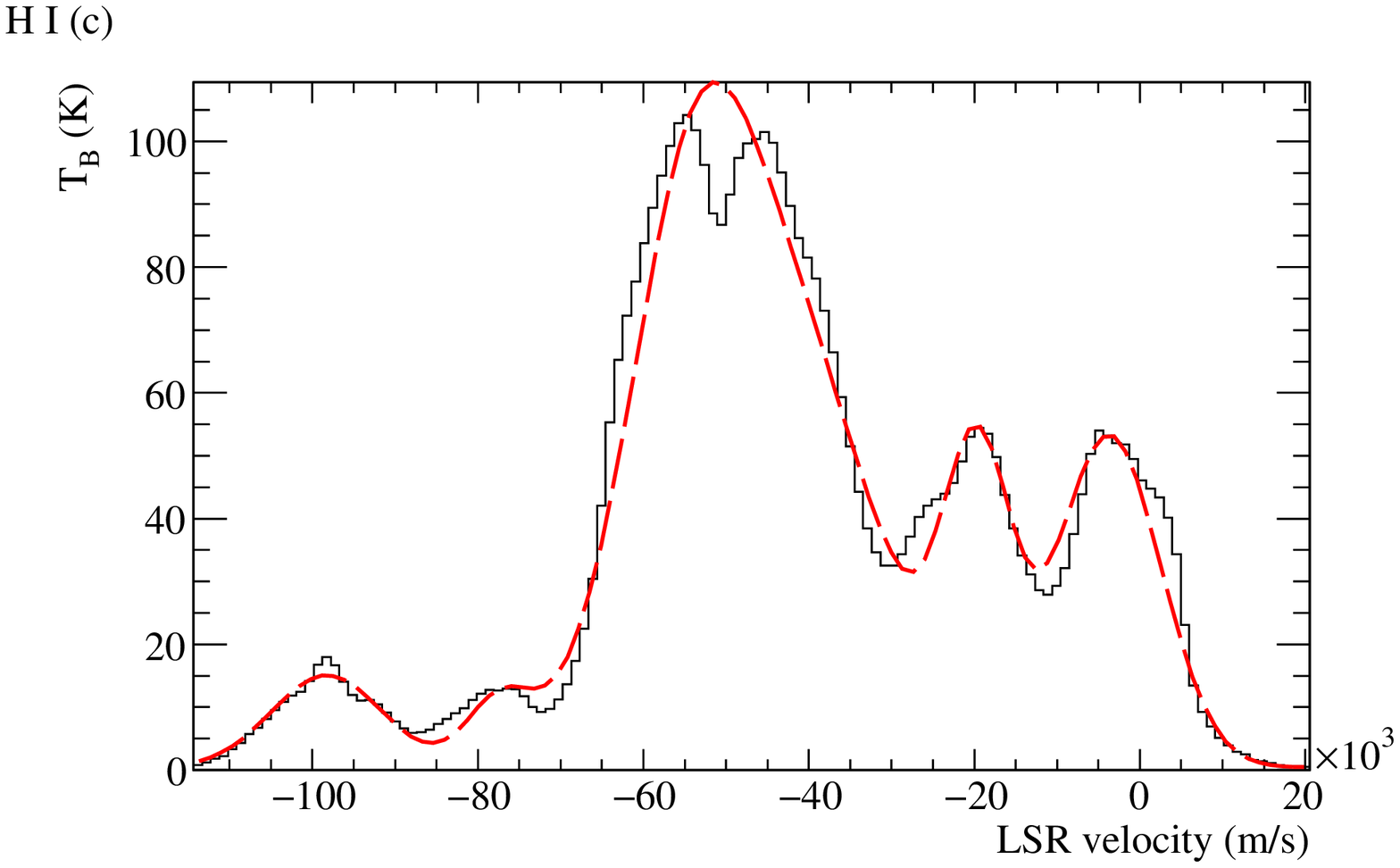} &
\end{tabular}
\end{center}
\caption{An example of the separation procedure described in~\ref{separation}
for the direction $l=133^\circ$ $b=0^\circ$. Each plot shows the brightness
temperature $T_{B}$ for $\hi$ or CO as a function of LSR velocity. Vertical
lines correspond to the boundaries Gould Belt -- local arm, local arm -- Perseus
arm, Perseus arm -- outer arm (from right to left). The three rows correspond
to: a) preliminary ring-boundaries, b) ``physical'' boundaries, c) Gaussian
fitting of the $\hi$ line.}\label{lines}
\end{figure*}

The separation of the structures along the
line of sight was achieved through a 3-step procedure:
\begin{itemize}
 \item[a)] preliminary separation based on Galactocentric rings;
 \item[b)] transformation of the ring-velocity boundaries into ``physical'' boundaries based on the $(l,b,v)$ coherence of clouds, and production of $\nhi$ and $\wco$ maps;
 \item[c)] correction of the $\nhi$ maps for the spill-over between adjacent regions.
\end{itemize}
The three steps are described in detail below. In Fig.~\ref{lines} the procedure
is illustrated for an example direction at $l=133^\circ$ $b=0^\circ$.

The preparation of the gas maps started from preliminary velocity boundaries given in terms of Galactocentric rings that roughly encompass the Gould Belt for $R<8.8$ kpc, the main part of the local arm at $8.8 $ kpc $ < R < 10$ kpc, the Perseus arm at $10 $ kpc $ < R < 14$ kpc, and the outer arm for $R>14$ kpc.  Following IAU recommendations, we adopted a flat rotation curve with $R_\sun=8.5$ kpc and a rotation velocity of 220 km s$^{-1}$ at the solar circle. The confusion that is apparent in the longitude-velocity $(l,v)$ diagram of Fig.~\ref{lvdiagram} between the Gould-Belt and local-arm components results from the integration over latitude and is much reduced in the actual $(l,b,v)$ cube which is used to construct the maps. The presence of two different components is evident in the example direction of Fig.~\ref{lines}: the first component peaks at $v \sim 0$ km~s$^{-1}$ (Gould Belt), the second one at $v \sim -15$ km~s$^{-1}$ (local arm).

Starting from this preliminary separation, the ring-velocity boundaries were adjusted for each line of sight to better separate structures on the basis of their coherence in the $(l,b,v)$ phase space. For each line of sight, every boundary was moved to the nearest minimum in the $\hi$ spectrum, or, if a minimum was not found, to the nearest saddle. The shifts are typically of the order of $1-10$ km~s$^{-1}$ (see Fig.~\ref{lines}~b). The adjusted boundaries were used to calculate $\nhi$ and $\wco$ in each region.

The broad $\hi$ clouds can easily spill over from one velocity interval into the next. To correct for this cross-contamination between adjacent intervals, for each line of sight the $\hi$ spectrum has been fitted by a combination of Gaussians (see Fig.~\ref{lines}~c). The overlap estimated from the fit was used to correct the column density $\nhi$ calculated in a specific interval from the spill-over from the adjacent regions. The correction on $\nhi$ is typically of the order of $1\%-10\%$, although it can reach $20\%-30\%$ in regions corresponding to the frontier between clouds in the Gould Belt and in the main part of the local arm.

\begin{figure*}[!p]
\begin{center}
\begin{tabular}{cc}
\includegraphics[width=0.45\textwidth]{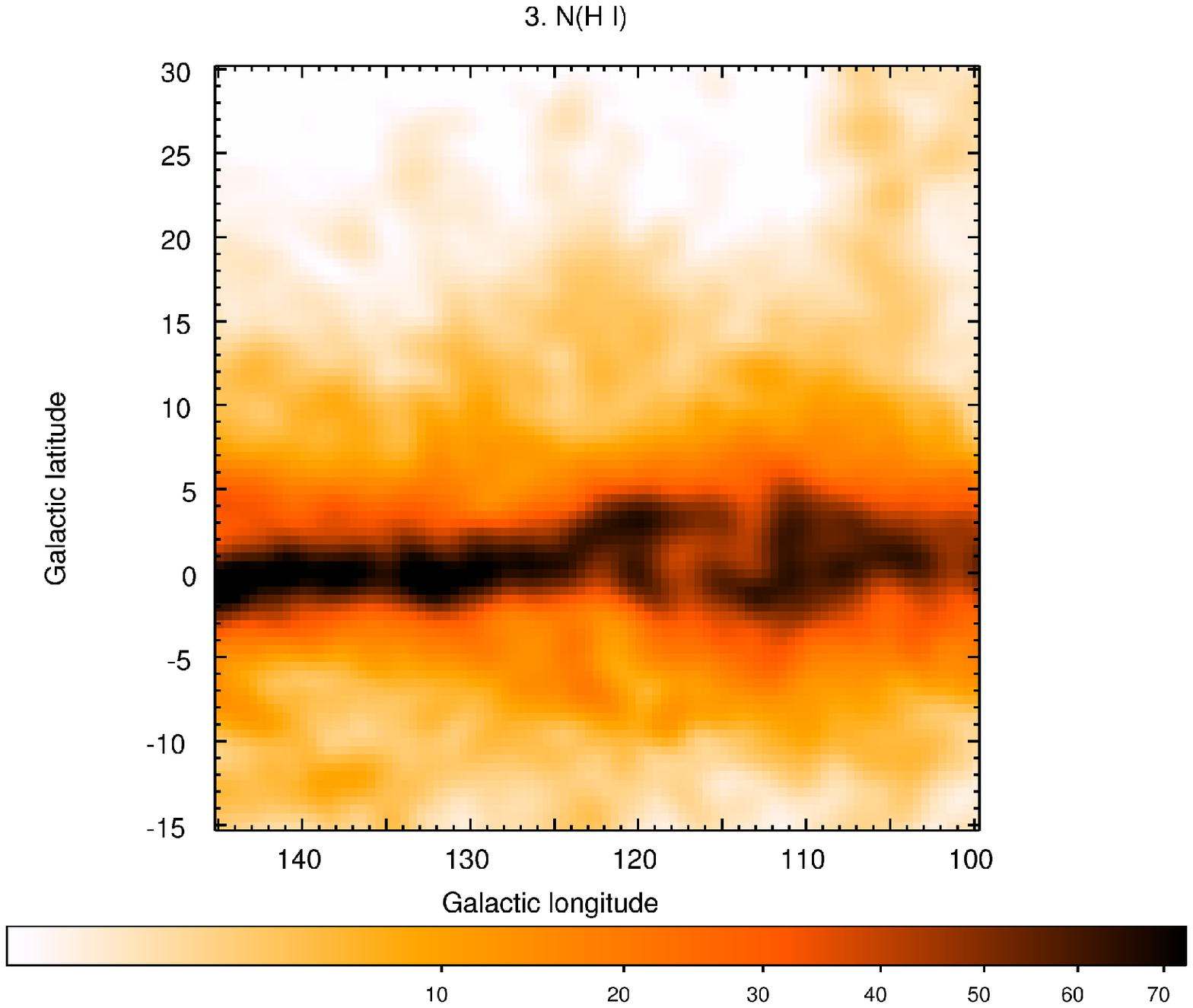} &
\includegraphics[width=0.45\textwidth]{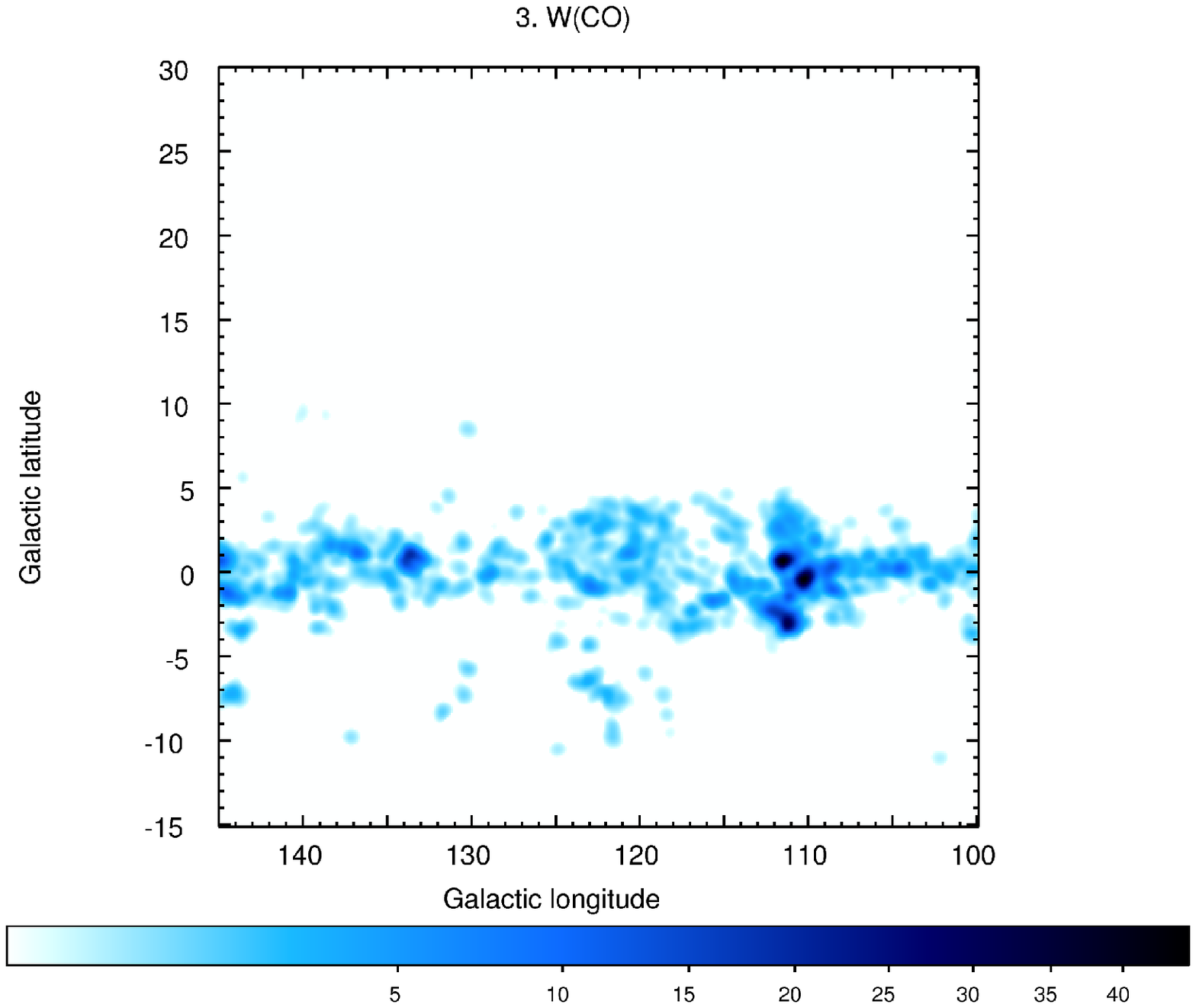} \\
\includegraphics[width=0.45\textwidth]{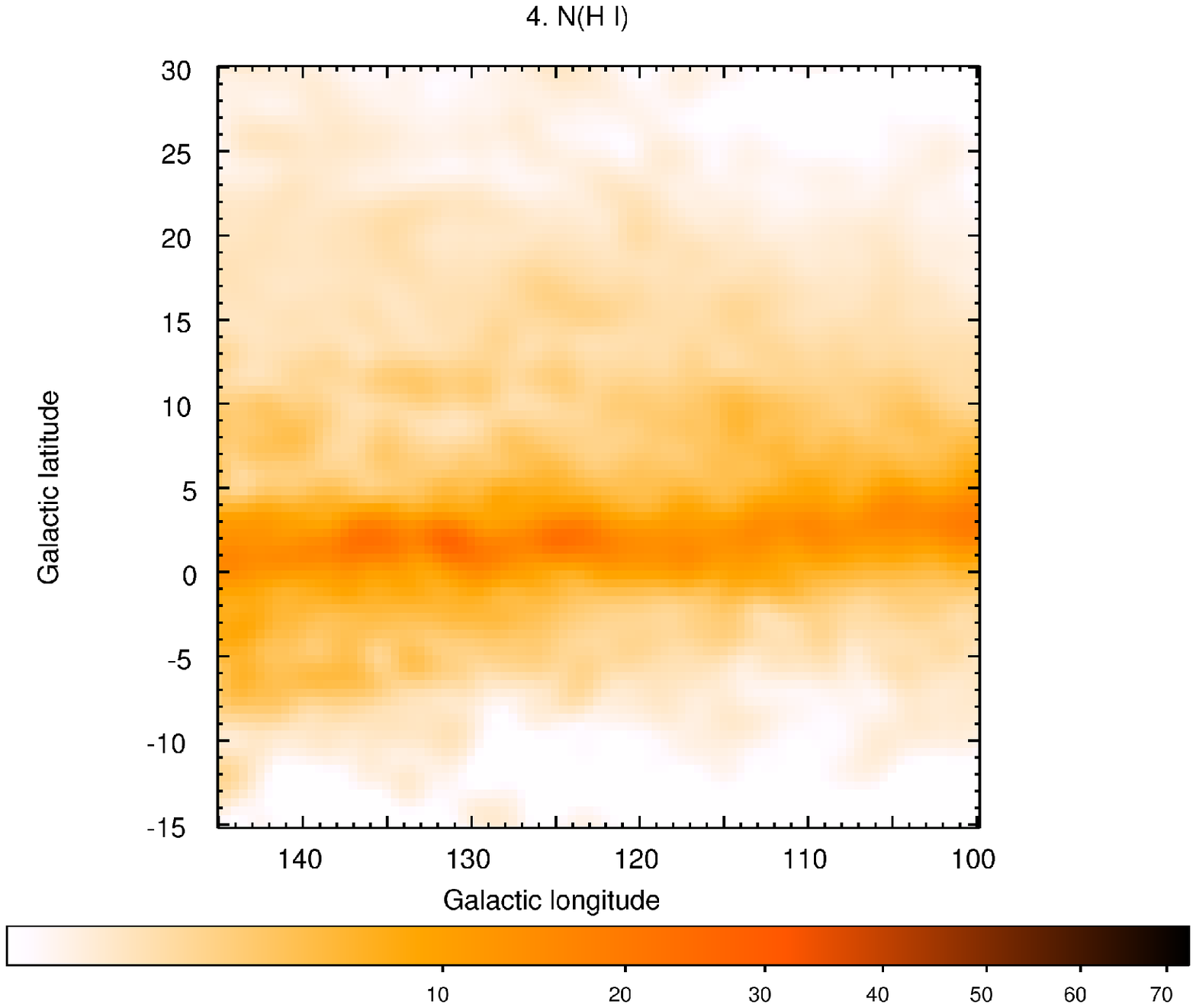} & \\
\end{tabular}
\end{center}
\caption{Maps of $\nhi$ (units of $10^{20}$ atoms cm$^{-2}$) and $\wco$ (units
of K km s$^{-1}$). Regions: (1) Gould Belt, (2) local arm, (3) Perseus arm, and
(4) outer arm. The maps have been smoothed for display with a Gaussian with
$\sigma=1^\circ$. Data sources are described in the
text.}\label{maps}
\end{figure*}

This separation scheme provides more accurate estimates of the actual gas mass
in a specific region and helps with separating structures. The resulting maps
are shown in Figure \ref{maps}.
\begin{figure*}[!p]
\begin{center}
\begin{tabular}{cc}
\includegraphics[width=0.45\textwidth]{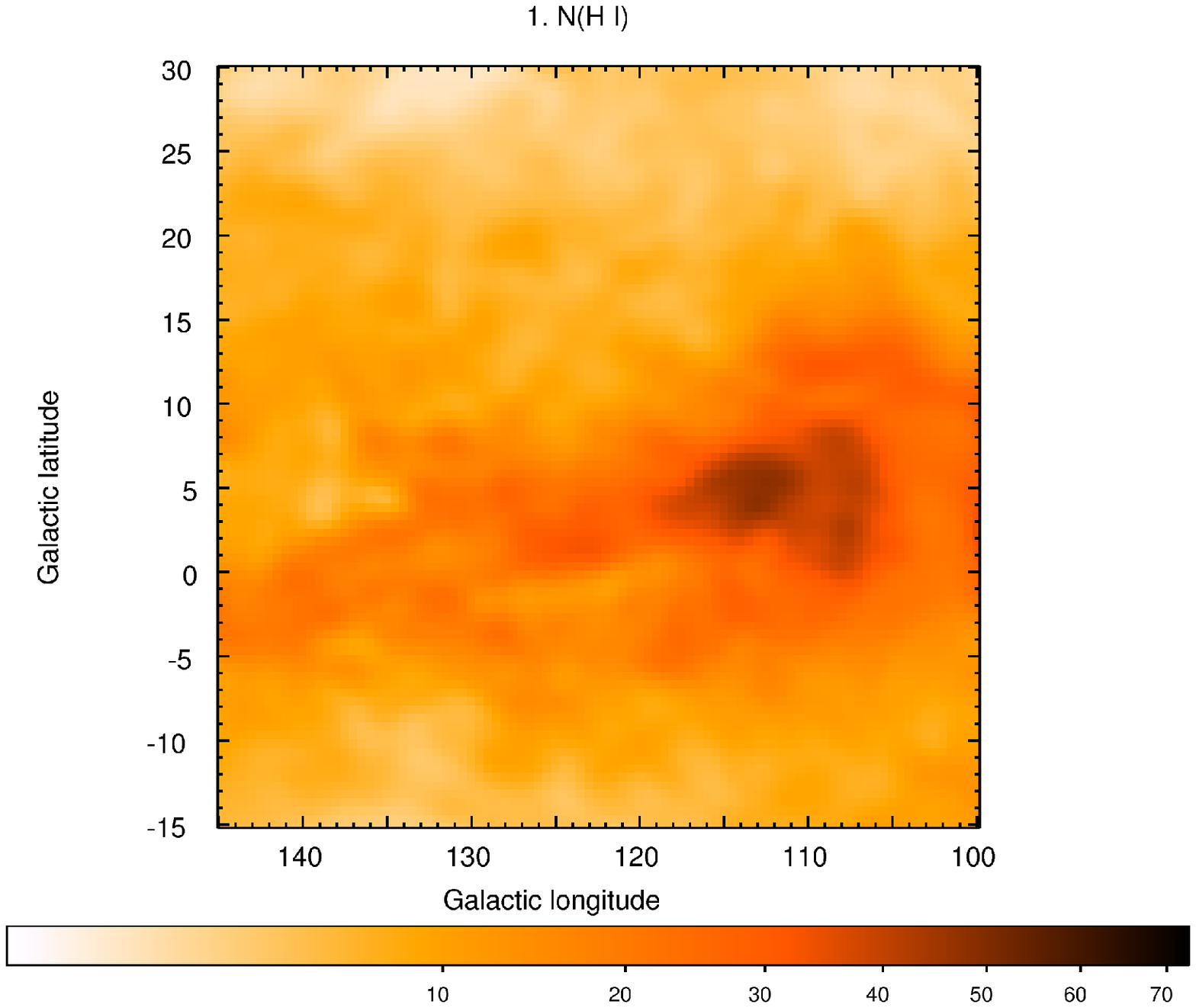} &
\includegraphics[width=0.45\textwidth]{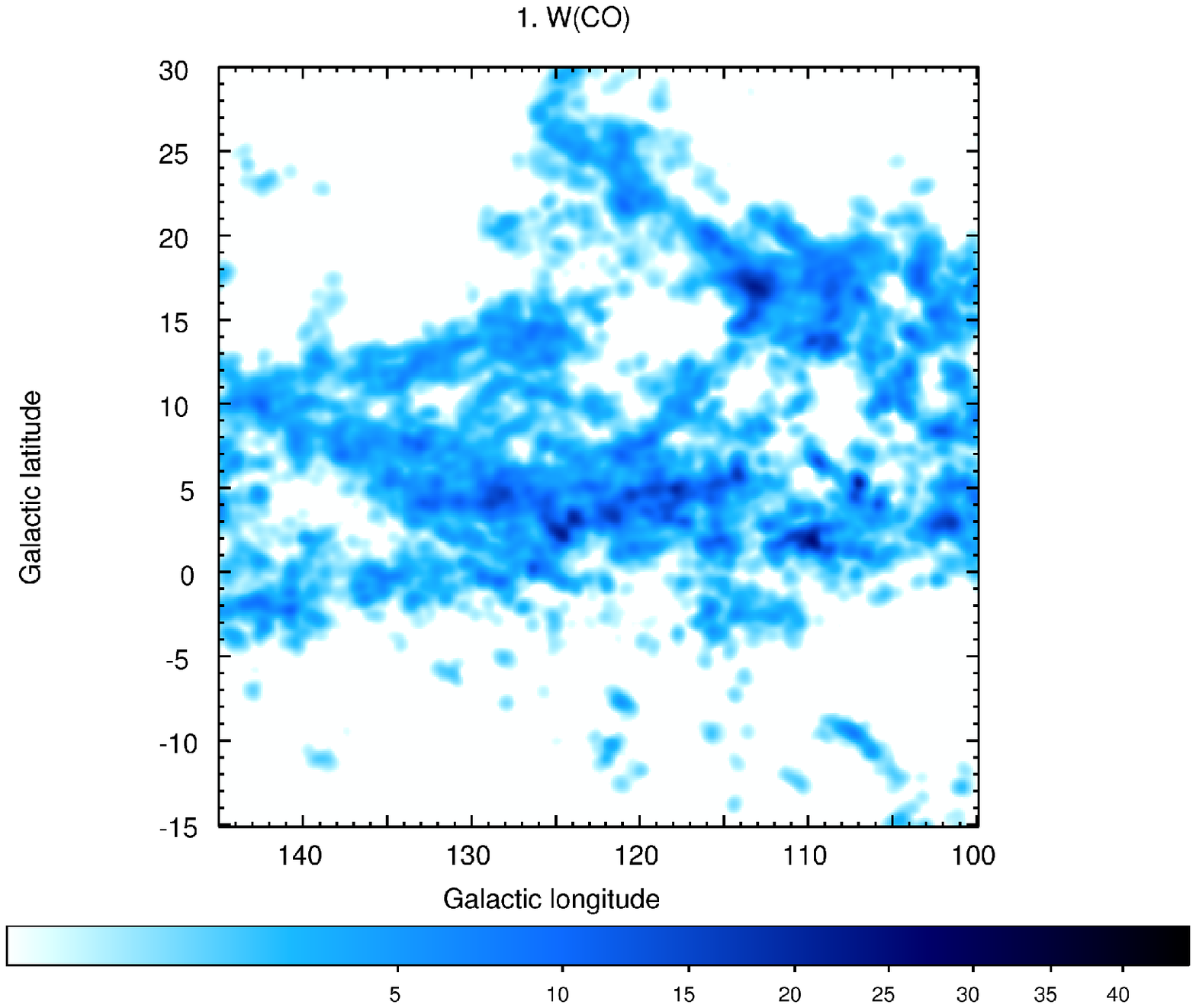}\\
\includegraphics[width=0.45\textwidth]{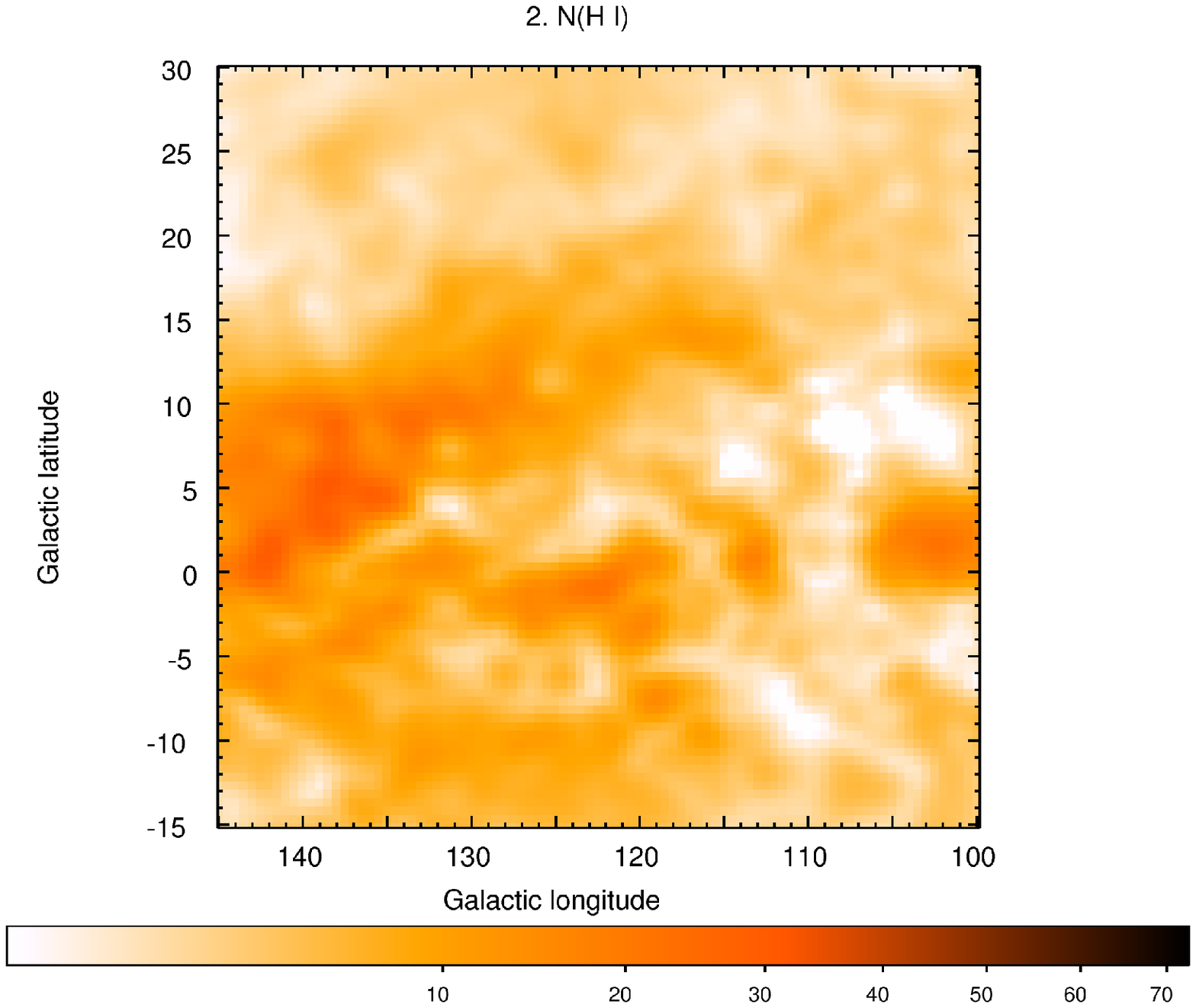} &
\includegraphics[width=0.45\textwidth]{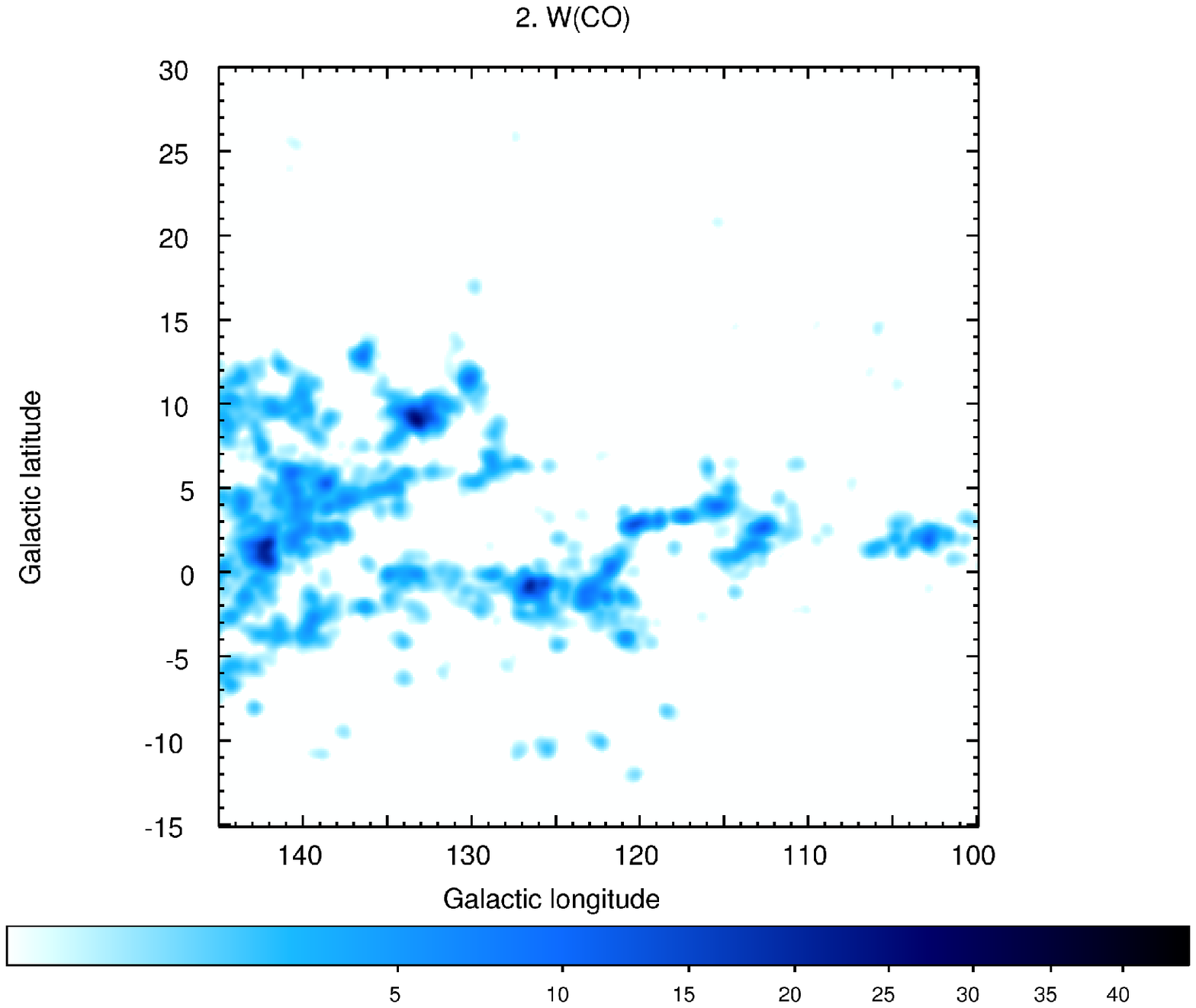} \\
\end{tabular}
\end{center}
\end{figure*}
They exhibit a low level of spatial degeneracy
between the cloud complexes found in the four regions along these directions.
Hence, we can model the observed $\gamma$-ray flux as a combination of
contributions coming from CR interactions in the different regions. The
correlation between the $\hi$ and CO phases in each region is unavoidable, but
not tight enough to hamper the separation between the $\gamma$-ray emission from
the two phases. No significant CO emission is found in the outer-arm region, so
the corresponding map was removed from the analysis. 

\subsection{Interstellar reddening}\label{dgsec}

An excess of $\gamma$ rays (observed by EGRET) correlated with an excess of dust thermal emission was found over the $\nhi$ and $\wco$ column-density maps in all the nearby Gould-Belt clouds by \citet{darkclouds}. Therefore, they reported a considerable amount of ``dark'' gas, i.e. neutral gas not properly traced by $\hi$ and CO, at the interface between the two radio-traced phases. The chemical state of the additional gas has not been determined yet, leaving room for $\hd$ poorly mixed with CO or to $\hi$, overlooked e.g. because of incorrect assumptions about the spin temperature for optical depth corrections or $\hi$ self absorption (see~\ref{hidata}).

Following the method proposed by \citet{darkclouds}, we have prepared a map to account for the additional gas. The map is derived from the $\ebv$ reddening map of \citet{ebv}, which provides an estimate of the total dust column densities across the sky. Point sources (corresponding to IRAS point sources) were removed and the corresponding pixels were set to the average value of the adjacent directions. In order to subtract the dust components correlated with $\nhi$ and $\wco$, the reddening map was fitted with a linear combination of the same set of $\nhi$ and $\wco$ maps for the Gould-Belt and local, Perseus, and outer-arm regions described above. A detailed discussion of the results of the fit goes beyond the scope of the present work, so it is deferred to another paper (A.~A.~Abdo et al. 2009, in preparation), which will address the results over several interstellar complexes in the Gould Belt and will compare them with $\gamma$-ray measurements by the LAT. 

\begin{figure}[!hbt]
\plotone{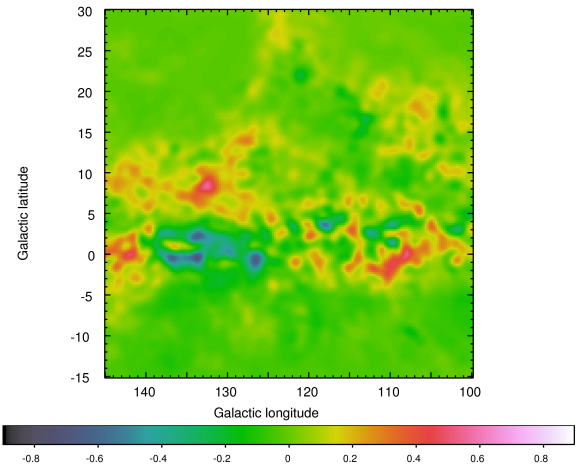}
\caption{$\ebvres$ map: map of the reddening residuals obtained after
subtraction of the parts linearly correlated with the combination of $\nhi$
column densities and $\wco$ intensities found in the four regions along the line
of sight (Gould Belt, local, Perseus, and outer arms). The positive residuals
surrounding CO clouds off the plane outline the potential dark gas envelopes of
the Gould Belt clouds. The map has been smoothed for display with a Gaussian
with $\sigma=1^\circ$.}\label{dgmap}
\end{figure}

The resulting $\ebvres$ residual map, obtained subtracting from the $\ebv$ map
the best-fit linear combination of our set of $\nhi$ and $\wco$ maps, is shown
in Fig.~\ref{dgmap}.

The residuals typically range from $-1$ to $+1$ magnitude.
Unlike in \citet{darkclouds}, both positive and negative residuals have been
considered in the analysis of the $\gamma$-ray data. Residuals hint at
limitations in the gas radio tracers as well as in the $\ebv$ map. Positive
residuals can correspond to a local increase in the dust-to-gas ratio and/or to
the presence of additional gas not properly accounted for in the $\nhi$ and
$\wco$ maps. The latter explanation is supported by the significant correlation
we will find between the $\ebvres$ map and the LAT $\gamma$-ray data
(see~\ref{dgdiscuss}). Fig.~\ref{dgmap} shows that at $|b|>5^\circ$ the
$\ebvres$ map is dominated by positive residuals forming structured envelopes
around the CO clouds. Small negative residuals are systematically seen toward
the CO cores. They may be due to a decrease in dust temperature in the denser,
well shielded, parts of the molecular complexes, or to local variations of the
dust-to-gas ratio. Further comparison with dust extinction tracers is needed to
investigate this effect. Positive and negative residuals appear at low latitude,
but, because of the pile up of dust clouds with different temperatures along the
line of sight, the temperature correction, applied by  \citet{ebv} to the
thermal emission to produce the $\ebv$ map, is not as valid near the plane as in
well resolved local clouds off the plane, in particular toward bright
star-forming regions. These effects may cause the clusters of negative residuals
at $|b|<5^\circ$. We note that the positive residuals along the plane are not
well correlated with the amount of self absorbed $\hi$ found in the second
quadrant \citep{gibson}. The most conspicuous self-absorbed $\hi$ cloud in fact
corresponds to the negative residuals seen at $130^\circ < l < 140^\circ$.

By construction, the $\ebvres$ map compensates for the limitations of the radio tracers, both by revealing non-emitting gas and by correcting the approximations applied to handle the radiative transfer of the radio lines. In particular the $\ebvres$ map depends on the optical depth correction applied to the $\nhi$ maps. We note that different choices of the $\hi$ spin temperature, from the optically thin case down to 250 and 125 K, change the $\ebv$ residuals by up to $0.2-0.3$ mag on the plane and 0.1 mag at $|b|>5^\circ$. Off the plane, the shape of the large structures of positive residuals around the Gould-Belt clouds hardly changes.

The strong correlation between $\ebv$ residuals and $\gamma$-ray data, which will be shown in~\ref{dgdiscuss}, proves that the interstellar reddening is in many directions a better tracer of the total gas column densities than the combination of $\hi$ and CO. Therefore, we will use the $\ebvres$ map to correct the standard radio and microwave tracers, very suitable for the aims of this work since, unlike reddening, they carry distance information.

\section{Gamma-ray analysis}

\subsection{LAT data}
The \emph{Fermi} LAT is a pair-tracking telescope \citep{latpaper}, detecting photons from $\sim 20$ MeV to more than 300 GeV. The tracker has 18 $(x,y)$ layers of silicon microstrip detectors interleaved with tungsten foils to promote the conversion of $\gamma$ rays into electron-positron pairs (12 thin foils of 0.03 radiation lengths in the front section plus 4 thick foils of 0.18 radiation lengths in the back section; the last two layers have no conversion foils). The tracker is followed by a segmented CsI calorimeter to determine the $\gamma$-ray energy. The whole system is surrounded by a scintillator shield to discriminate the charged cosmic-ray background. The instrument design and the analysis result in a peak effective area of $\sim 8000$ cm$^2$ ($\sim$ 6 times greater than EGRET), a field of view of $\sim 2.4$ sr ($\sim 5$ times greater than EGRET) and a superior single photon angular resolution (for front converting photons, the 68\% containment angle at 1 GeV reaches $\sim 0.6^\circ$ with respect to $\sim 1.7^\circ$ for EGRET).

Data were obtained during the period 2008 August 4 - 2009 July 4. The \emph{Fermi} observatory was operated in scanning sky survey mode, rocking $35^\circ$ north and south of the zenith on alternate orbits, apart from calibration runs that are excluded from the analysis. We used the dataset prepared for the construction of the first year Catalog of LAT sources (A.~A.~Abdo et al. 2009, in preparation), excluding brief time intervals corresponding to bright $\gamma$-ray bursts. It uses the \emph{Diffuse} event selection, which has the least residual CR background contamination \citep{latpaper}. We also selected events on the basis of the measured zenith angle to limit the contamination from interactions of cosmic rays with the upper atmosphere of the Earth. Owing to these interactions the limb of the Earth is a very bright $\gamma$-ray source, seen at a zenith angle of $\sim 113^\circ$ at the 565 km nearly-circular orbit of \emph{Fermi}. Since our region is close to the North celestial pole it is often observed at large rocking angles. In order to reduce the Earth albedo contamination, we accept for analysis here only events seen at a zenith angle $<100^\circ$. The exposure is only marginally affected (because the detection efficiency dramatically decreases at large inclination angles), but the background rate is significantly reduced.

\subsection{Model for analysis}\label{anaproc}

The analysis scheme used since the COS-B era \citep{lebrun,gradient,egretcep} is
based on a very simple transport model. Assuming that the interstellar medium is
transparent to $\gamma$ rays, that the characteristic diffusion lengths for CR
electrons and protons exceed the dimensions of cloud complexes, and that cosmic
rays penetrate clouds uniformly to their cores, the $\gamma$-ray intensity $I$
(cm$^{-2}$ s$^{-1}$ sr$^{-1}$) in a direction $(l,b)$ can be modeled to first
order as a linear combination of contributions coming from CR interactions with
the different gas phases in the various regions along the line of sight. We add
the contribution from point-like sources and an isotropic intensity term.
Several processes are expected to contribute to the latter, notably the
extragalactic $\gamma$-ray background and the residual instrumental background
from misclassified interactions of charged CRs in the LAT. The IC emission is
also expected to be rather uniform across this small region of the sky. We used
the current best models of IC emission to verify that it is statistically not
distinguishable from an isotropic background over the small region of interest,
at large angular distance from the inner Galaxy (see \ref{param}). The present
analysis does not aim to provide meaningful results for the extragalactic
background and the IC emission which will be addressed in forthcoming
publications (\citealt{EGBpap}, A.~A.~Abdo et al. 2009, in preparation).

In the absence of suitable tracers for the diffuse ionized gas (primarily $\hii$), the derived $\gamma$-ray emissivities for neutral gas will be slightly overestimated. However, the ionized gas is contributing to $\sim 10\%$ of the total mass and, because of its large scale height of $\sim 1$ kpc above the plane \citep{hiimodel}, part of its $\gamma$-ray emission will be overtaken by the isotropic term in the fit to the LAT data. So the bias on the neutral gas emissivities should be small.

Therefore, the $\gamma$-ray intensity $I$, integrated in a given energy band, is modeled by Eq.~\ref{anamodel}.
\begin{eqnarray}
\hspace{-2cm} I(l,b) & = & \sum_{\imath=1}^4 \left[ \qhi{\imath} \cdot
\nhi(l,b)_\imath + \qco{\imath} \cdot \wco(l,b)_\imath \right] +\nonumber \\
&+& \qebv \cdot \ebvres(l,b) + I_\mathrm{iso} + \nonumber \\
&+& \sum_{\jmath} S_\jmath \cdot \delta^{(2)}(l-l_\jmath,b-b_\jmath)\label{anamodel}
\end{eqnarray}
The sum over $\imath$ represents the combination of the four Galactic regions. The free parameters are the emissivities of $\hi$ gas, $\qhi{\imath}$ (s$^{-1}$ sr$^{-1}$), per unit of $\wco$ intensity, $\qco{\imath}$ (cm$^{-2}$ s$^{-1}$ sr$^{-1}$ (K km s$^{-1}$)$^{-1}$), and per unit of $\ebv$ residuals, $\qebv$ (cm$^{-2}$ s$^{-1}$ sr$^{-1}$ mag$^{-1}$). $I_\mathrm{iso}$ (cm$^{-2}$ s$^{-1}$ sr$^{-1}$) is the isotropic background intensity. The contribution from point sources is represented by the sum over $\jmath$, where $S_\jmath$ is the integrated flux (cm$^{-2}$ s$^{-1}$) of the source lying at the position $(l_\jmath,b_\jmath)$.

\subsection{Analysis procedure}

\subsubsection{Method}

We used the standard LAT analysis environment provided by the \emph{Science
Tools}\footnote{\url{
http://fermi.gsfc.nasa.gov/ssc/data/analysis/documentation/Cicerone/}}. The
$\gamma$-ray statistics are large enough to model the spectral shape of each
component as a power law in relatively narrow energy bands. This assumption,
together with the iterative procedure described below in \ref{anastep}, allows
the exposures and the convolution with the energy-dependent Point Spread
Function (PSF) to be computed without forcing an \emph{a~priori} spectral index.
The \emph{Science Tools} provide a full convolution of the maps with the
energy-dependent PSF. The \emph{Science Tools} are also very flexible in the
description of point sources (number, location, spectra). We used the
\texttt{P6\_V3} post launch Instrument Response Functions (IRFs), which take
into account the loss of detection efficiency due to pile-up and accidental
coincidence effects in the LAT \citep{rando}.

LAT data have been analysed using a binned maximum-likelihood procedure with Poisson statistics, on a spatial grid with $0.5^\circ$ spacing in Cartesian projection. The higher energy range we have investigated starts at a few GeV, where the 68\% containment angle is $\sim 0.5^\circ$ for events converting in the front section of the tracker (about a factor two larger for back converting events), so we cannot resolve details smaller than this in the $\gamma$-ray maps. This resolution is commensurate with that of the $\hi$ and $\ebv$ maps.

\begin{figure*}[!p]
\begin{tabular}{cc}
\includegraphics[width=0.45\textwidth]{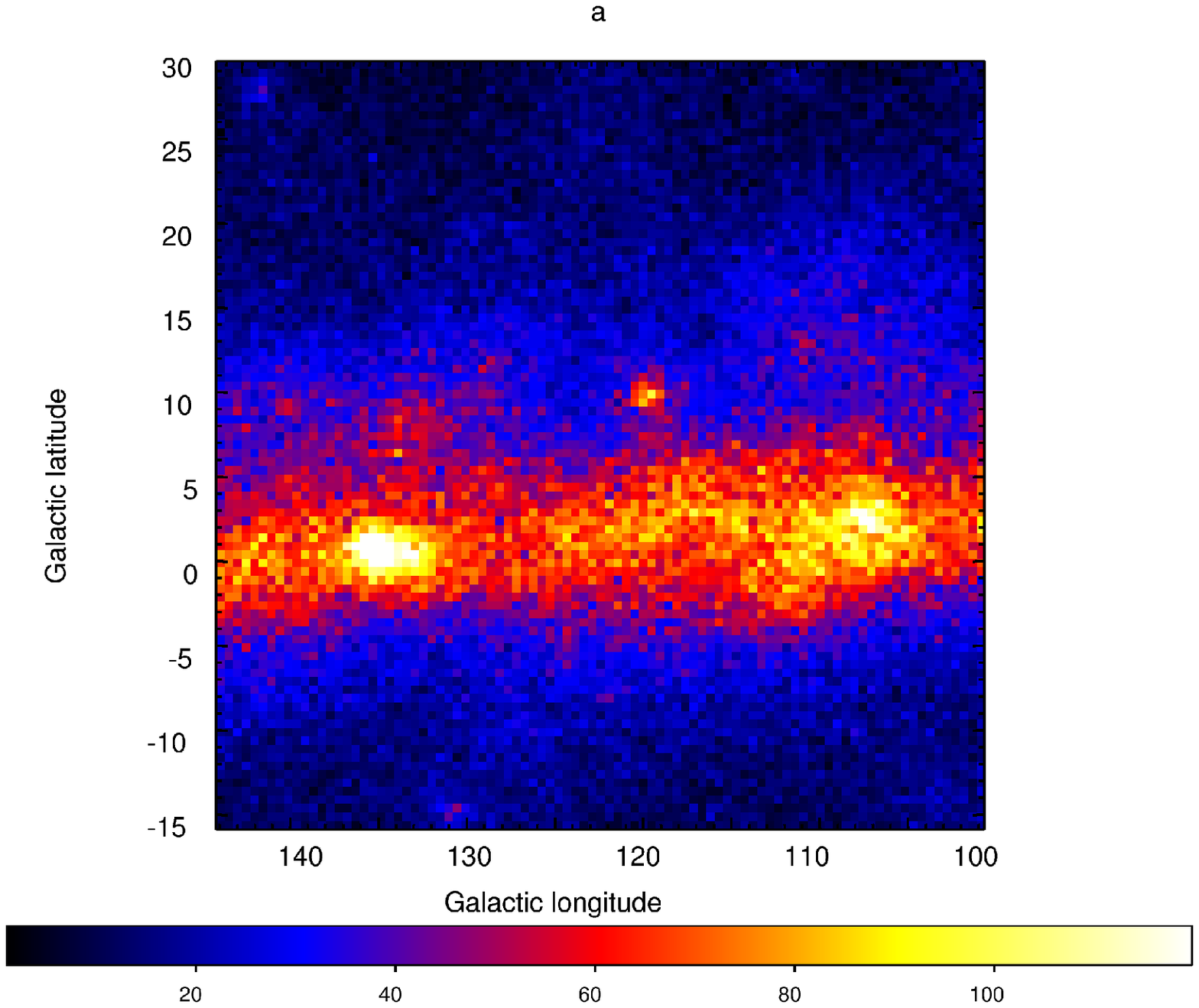}&\includegraphics[
width=0.45\textwidth]{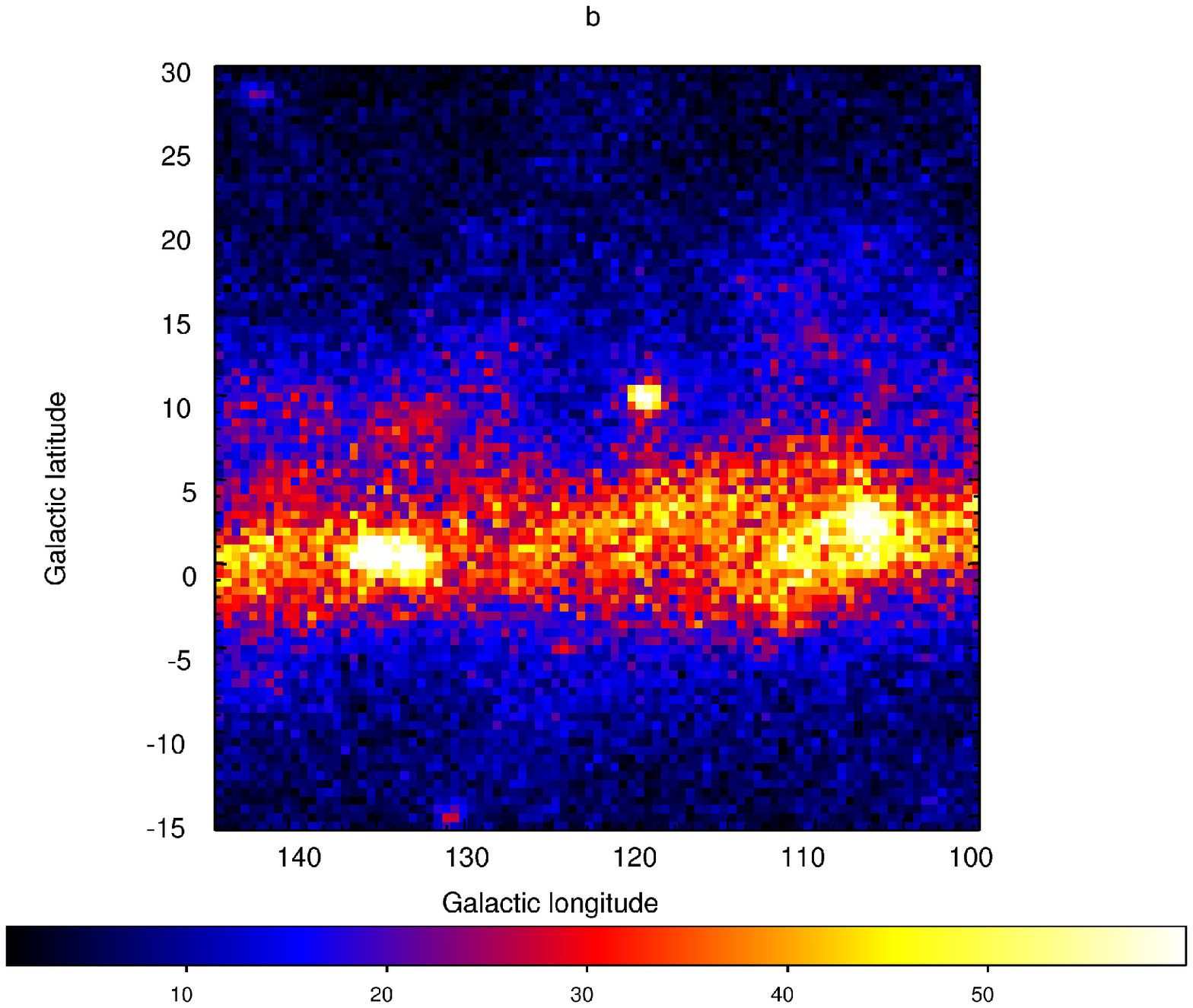}\\
\includegraphics[width=0.45\textwidth]{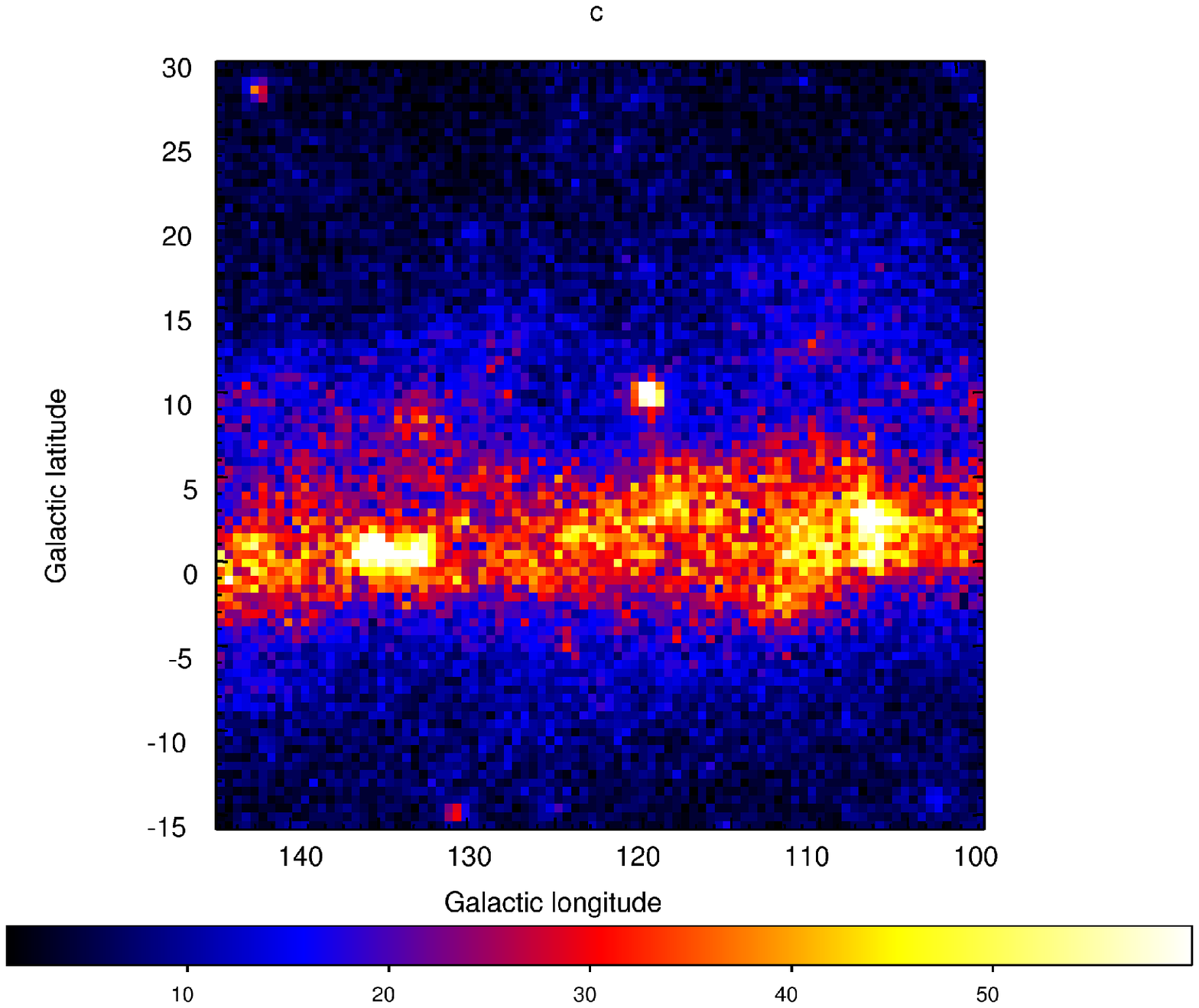}&\includegraphics[
width=0.45\textwidth]{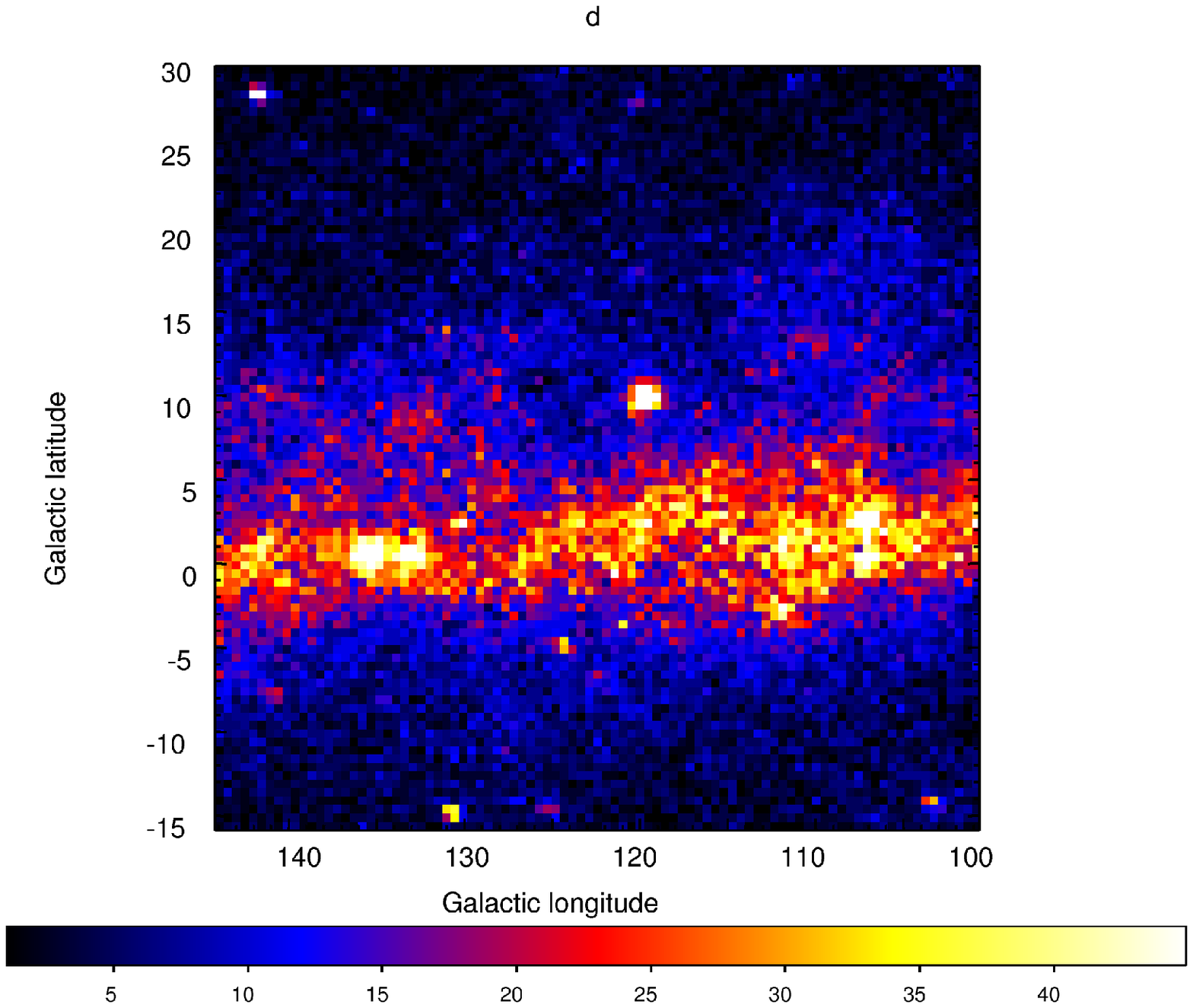}\\
\includegraphics[width=0.45\textwidth]{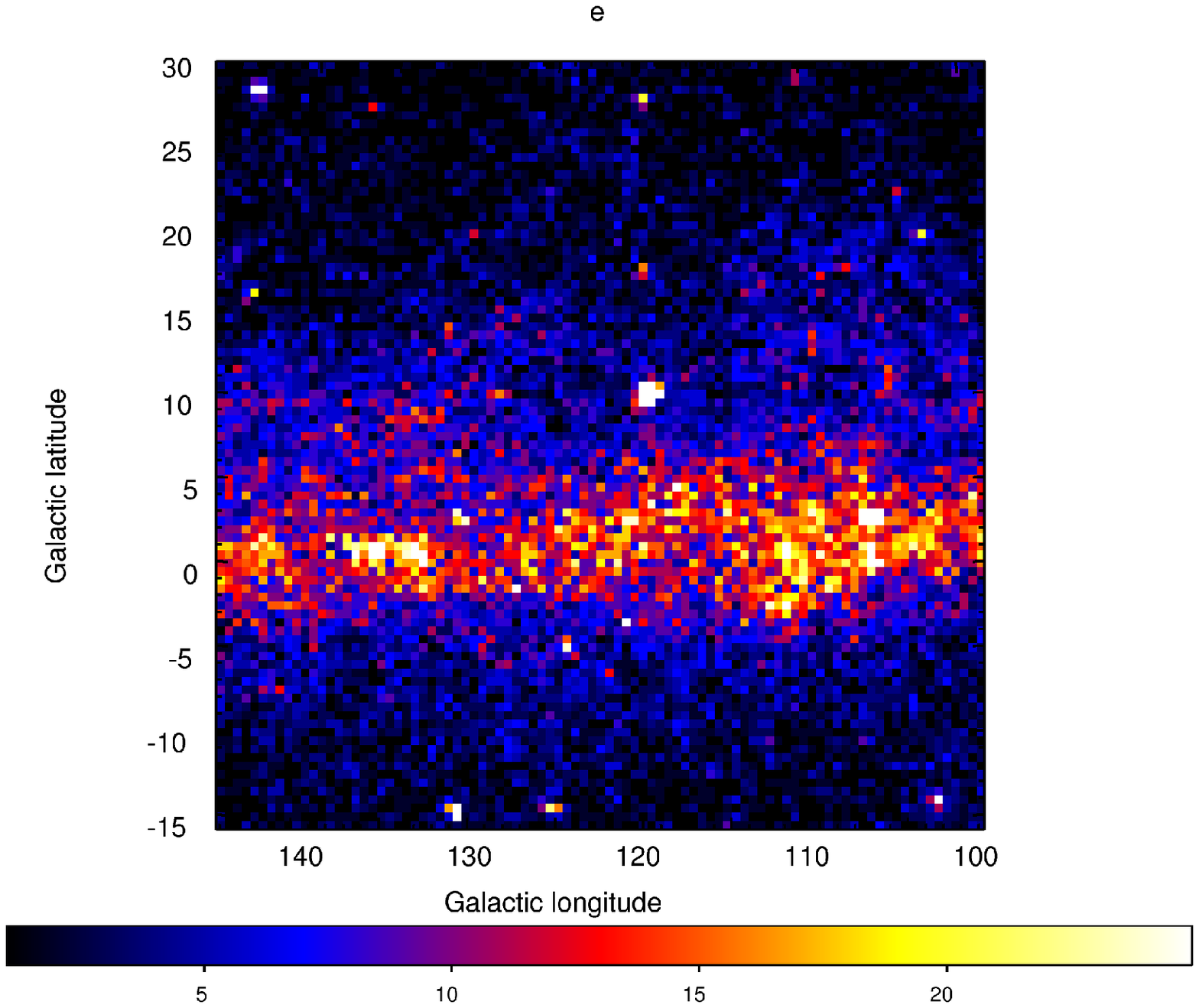}
\end{tabular}
\caption{Gamma-ray count maps in the five energy bands: a) 200 MeV -- 400 MeV,
b) 400 MeV -- 600 MeV, c) 600 MeV -- 1 GeV, d) 1 -- 2 GeV, e) 2 -- 10
GeV.}\label{cmaps}
\end{figure*}

The analysis was performed for 5 contiguous energy bands: 200 MeV -- 400 MeV,
400 MeV -- 600 MeV, 600 MeV -- 1 GeV, 1 GeV -- 2 GeV and 2 GeV -- 10 GeV. The
energy bands were chosen wide enough to obtain stable results for the fit
parameters, because large statistical fluctuations might hamper the separation
of the different maps. Below 200 MeV the broad PSF does not allow an effective
separation of the different maps. We are confident that between 0.2 and 10 GeV
the interstellar $\gamma$-ray emission from the gas dominates over the
instrumental foregrounds. The count maps in the five energy bands are shown in
Fig.~\ref{cmaps}.

\subsubsection{Point sources}\label{anastep}

The inclusion of sources in the analysis model is a non-trivial task because the likelihood maximization procedure (based on the optimization engine \texttt{Minuit2}\footnote{\url{http://wwwasdoc.web.cern.ch/wwwasdoc/minuit/minmain.html}}) is stable up to a few tens of free parameters. The sources have thus been added following an iterative procedure.

The sources were taken from the 11 month source list, which will be the basis for the First Year Catalog of LAT sources in preparation (A.~A.~Abdo et al. 2009, in preparation). The sources were added following the detection significance (TS) in the 11 month source list\footnote{The Test Statistic, TS, is defined as $\mathrm{TS}=2(\ln \like - \ln \like_0)$, where $\like$ and $\like_0$ are the maximum-likelihood values reached with and without the source, respectively.}. The sources are added as point-like sources keeping their positions at those given in the list while letting their power-law spectra to vary independently in each energy band. No further attempts are made in this analysis to improve the spectral modeling or to account for possible extension.

The inclusion of the sources went through the following steps, where the parameters of the diffuse emission model were always let free:
\begin{itemize}
 \item We started with no point sources in the model.
 \item We added 9 sources detected with $\mathrm{TS}>600$ (hereafter bright sources). They were added to the sky model 3 at a time in order of decreasing TS, freezing at each step the previous source spectra and fitting the last three, while the diffuse parameters were always let free. Among bright sources, for the 6 sources lying in the region under study, we let their fluxes and spectral indexes free; for 3 sources lying just outside ($< 5^\circ$) the region boundaries, we fixed their parameters at the values determined in the 11 month source list.  These bright sources were already reported in the LAT Bright Source List \citep{BSlist}: two of them are firmly identified as pulsars (0FGL~J$0007.4+7303$ or LAT~PSR~J0007$+$7303, and 0FGL~J$2229.0+6114$ or PSR~J2229$+$6114), one as a $\gamma$-ray binary (0FGL~J$0240.3+6113$ or LSI+61~303), and the others are associated with blazars.
 \item We then added 52 more sources in the 11 month source list within the region boundaries with $\mathrm{TS}$ between 600 and $\simeq 25$ (out of them 22 where detected with $\mathrm{TS}>100$); they were added in several groups of 6 or 5 sources, with a procedure analogous to that used to handle bright sources, but only their integrated fluxes were allowed to be free, whereas the spectral indexes were fixed at the values in the source list.
 \item Finally, the analysis was repeated with all the sources, letting free only the parameters of the diffuse model and of the bright sources.
\end{itemize}

The iterative procedure allows verifying that only the bright sources can affect the diffuse parameters: the latter do not significantly change when less significant sources ($\mathrm{TS} < 600$) are added to the model. This does not apply to $I_\mathrm{iso}$ and $\qhi{4}$: we note that their values keep decreasing as we add new sources down to $\mathrm{TS} \sim 25$. We argue that the isotropic intensity generally absorbs point sources off the plane that are not included in the analysis; as said before we are not trying to give a physical interpretation of $I_\mathrm{iso}$. On the other hand, given the low linear resolution in distant clouds of the outer arm and the subsequent lack of pronounced features in the map (see Fig.~\ref{maps}), point sources at very low latitude ($|b|\lesssim 3^\circ$) can strongly bias the value of the corresponding $\hi$ emissivity, $\qhi{4}$, as separating them from the clumpy ISM emission near the plane in $0.5^\circ$ maps is difficult. Therefore, we consider this parameter only as an upper limit to the real gas emissivity in the outer arm.

\subsection{Fit results}\label{param}
\begin{figure*}[!p]
\begin{tabular}{cc}
\includegraphics[width=0.45\textwidth]{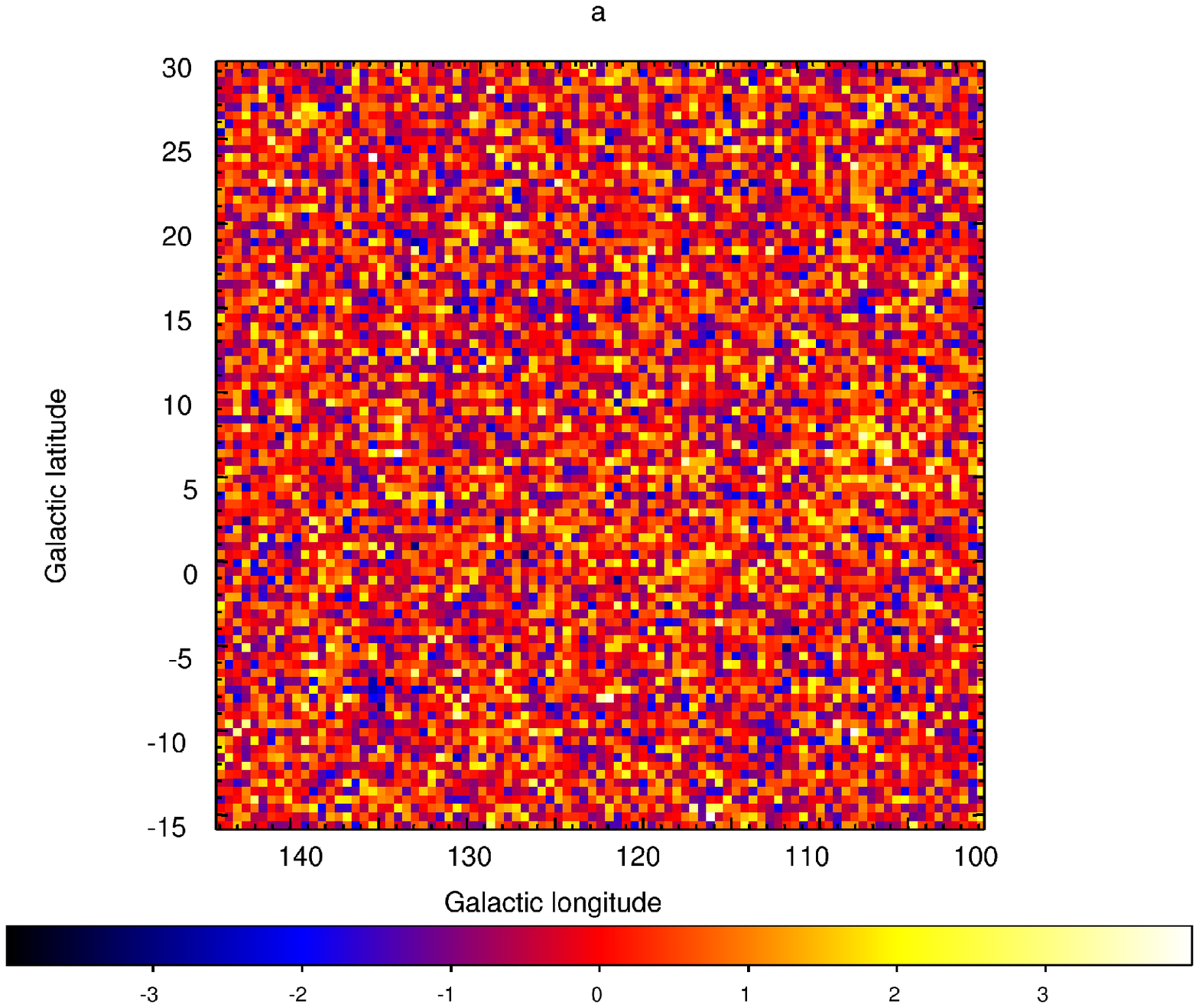}&\includegraphics[
width=0.45\textwidth]{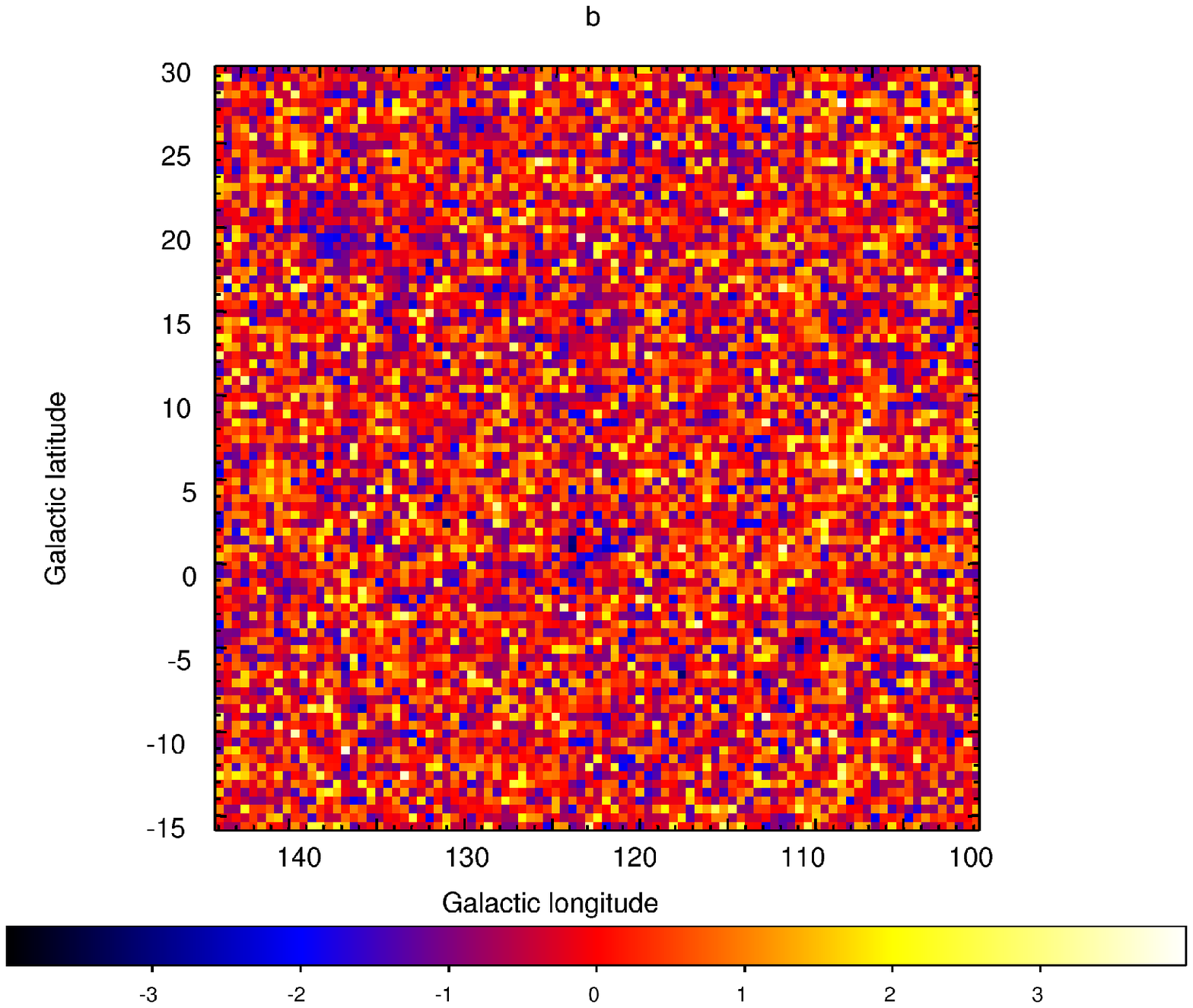}\\
\includegraphics[width=0.45\textwidth]{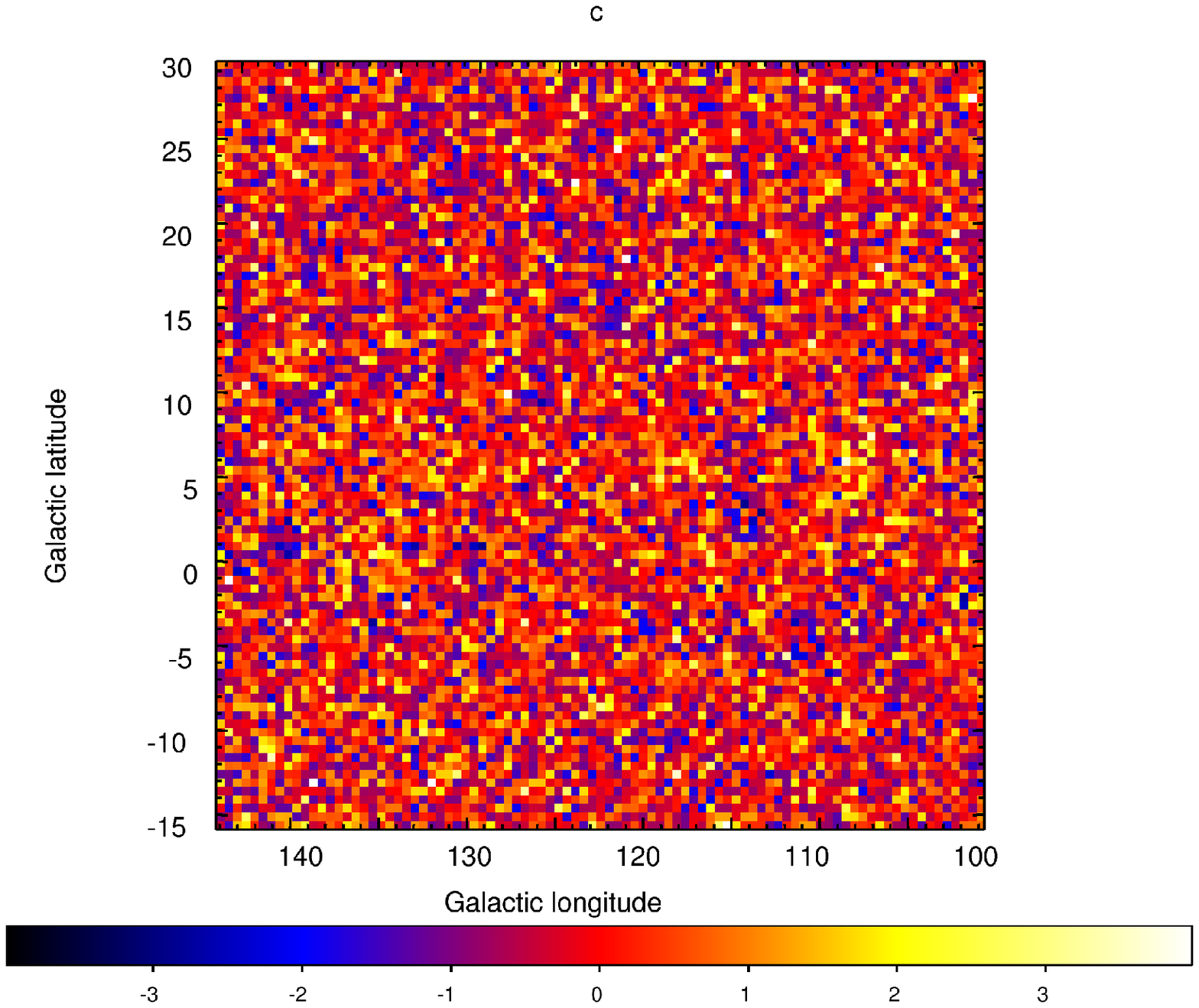}&\includegraphics[
width=0.45\textwidth]{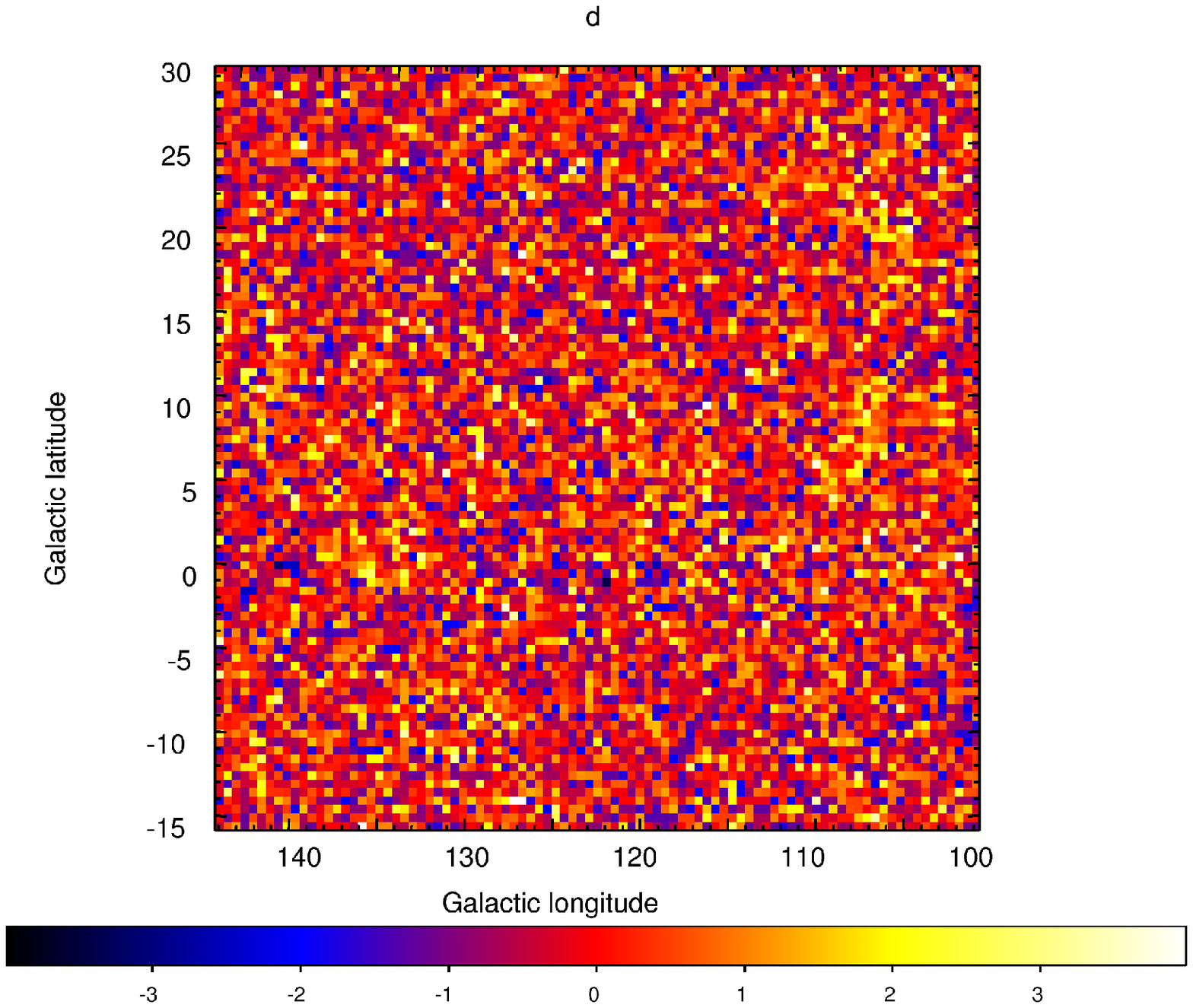}\\
\includegraphics[width=0.45\textwidth]{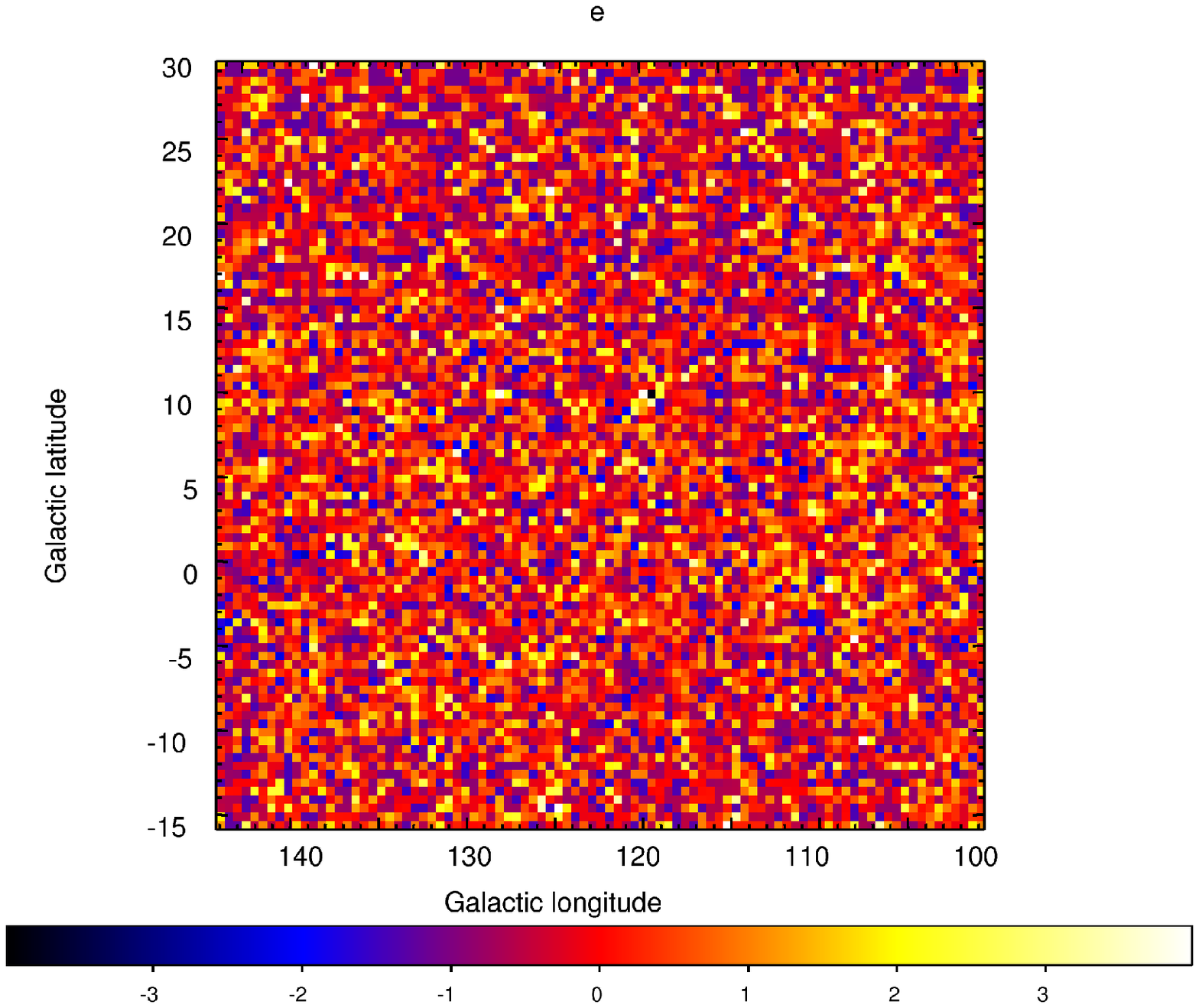}
\end{tabular}
\caption{Gamma-ray residual maps in the same energy bands as in
Fig.~\ref{cmaps}. The residuals, i.e. observed counts minus model-predicted
counts, are in units of the square root of the model-predicted counts (truncated
between $-4$ and $+4$ for display).}\label{sigmaps}
\end{figure*}

The quality of the final fits is illustrated in the residual maps of
Fig.~\ref{sigmaps}.
The residuals, i.e. observed counts minus model-predicted
counts, are expressed in standard deviation units (square root of
model-predicted counts). The maps show no excesses below $-4\sigma$ or above
$+7\sigma$.

\begin{deluxetable}{c r @{$\pm$} l r @{$\pm$} l r @{$\pm$} l r @{$\pm$} l r
@{$\pm$} l}
\tablewidth{0pt} 
\tablecaption{Parameters of the diffuse emission model obtained from the fit to
LAT data. \label{fitparameters}}
\tablecolumns{6}
\tablehead{
\colhead{parameter\tablenotemark{a,}\tablenotemark{b}} & \multicolumn{2}{c}{0.2
-- 0.4 GeV} & \multicolumn{2}{c}{0.4 -- 0.6 GeV} & \multicolumn{2}{c}{0.6 -- 1
GeV} & \multicolumn{2}{c}{1 -- 2 GeV} & \multicolumn{2}{c}{2 -- 10 GeV} 
}
\startdata
$\qhi{1}$		& 0.584&0.011	& 0.224&0.008	& 0.168&0.004 	&
0.110&0.003	& 0.048&0.002\\
$\qco{1}$		& 1.09&0.04	& 0.367&0.017 	& 0.318&0.013 	&
0.198&0.008	& 0.102&0.005\\
$\qhi{2}$		& 0.536&0.018	& 0.200&0.007 	& 0.157&0.005 	&
101&0.004	& 0.054&0.002\\
$\qco{2}$		& 1.67&0.17	& 0.47&0.06 	& 0.44&0.04 	&
0.26&0.03	& 0.087&0.014\\
$\qhi{3}$		& 0.349&0.011	& 0.128&0.004 	& 0.108&0.003 	&
0.072&0.002	& 0.0397&0.0014\\
$\qco{3}$		& 1.17&0.15	& 0.52&0.06 	& 0.37&0.04 	&
0.24&0.03	& 0.115&0.016\\
$\qhi{4}$		& 0.33&0.04	& 0.101&0.017 	& 0.114&0.013	&
0.103&0.009	& 0.032&0.005\\
$\qebv$			& 16.7&1.0	& 6.0&0.4 	& 3.49&0.27	&
2.28&0.18	& 0.80&0.11\\
$I_\mathrm{iso}$	& 4.67&0.10	& 1.19&0.04 	& 0.92&0.03	&
0.63&0.02	& 0.371&0.0017
\enddata
\tablenotetext{a}{Units: $\qhi{\imath}$ ($10^{-26}$ s$^{-1}$ sr$^{-1}$),
$\qco{\imath}$ ($10^{-6}$ cm$^{-2}$ s$^{-1}$ sr$^{-1}$ (K km s$^{-1}$)$^{-1}$),
$\qebv$ ($10^{-6}$ cm$^{-2}$ s$^{-1}$ sr$^{-1}$ mag$^{-1}$), $I_\mathrm{iso}$
($10^{-6}$ cm$^{-2}$ s$^{-1}$ sr$^{-1}$).}
\tablenotetext{b}{The subscripts refer to the different regions under analysis:
1) Gould Belt, 2) local arm, 3) Perseus arm, 4) outer arm and beyond.}
\end{deluxetable}

The best-fit parameters obtained in the five energy bands are given in
Table~\ref{fitparameters}, where the uncertainties correspond only to
statistical errors. We have also evaluated the systematic errors due to the
uncertainties on the event selection efficiency. From the comparison between
Monte Carlo simulations and real observations of the Vela pulsar, they are
evaluated to be 10\% at 100 MeV, 5\% at 500 MeV and 20\% at 10 GeV, scaling
linearly with the logarithm of energy between these values. These uncertainties
were parametrized into two sets of IRFs encompassing the most extreme scenarios.
The last step of the analysis has been repeated using these two IRF sets and the
results are assumed to bracket the systematic errors due to the event selection
efficiency (shown as shaded grey areas in the following figures). Only the last
step was considered, because we previously verified that only bright sources
impact the parameters of the diffuse emission model.

We also verified the impact of the isotropic approximation for the IC emission, repeating the last step of the analysis including a recent model based on the GALPROP CR propagation code \citep[see e.g.][]{galprop98,galprop04,porterIC}. The values obtained for the parameters of the diffuse emission model were compatible with the previous results, except for the isotropic intensity.

Other systematic uncertainties will be addressed in the discussion section.

\section{Discussion}

\subsection{Emissivity per $\mathbf{\hi}$ atom and cosmic-ray spectra}

\subsubsection{Consistency with other measurements}

\begin{figure*}[!bt]
\begin{tabular}{cc}
\includegraphics[width=0.48\textwidth]{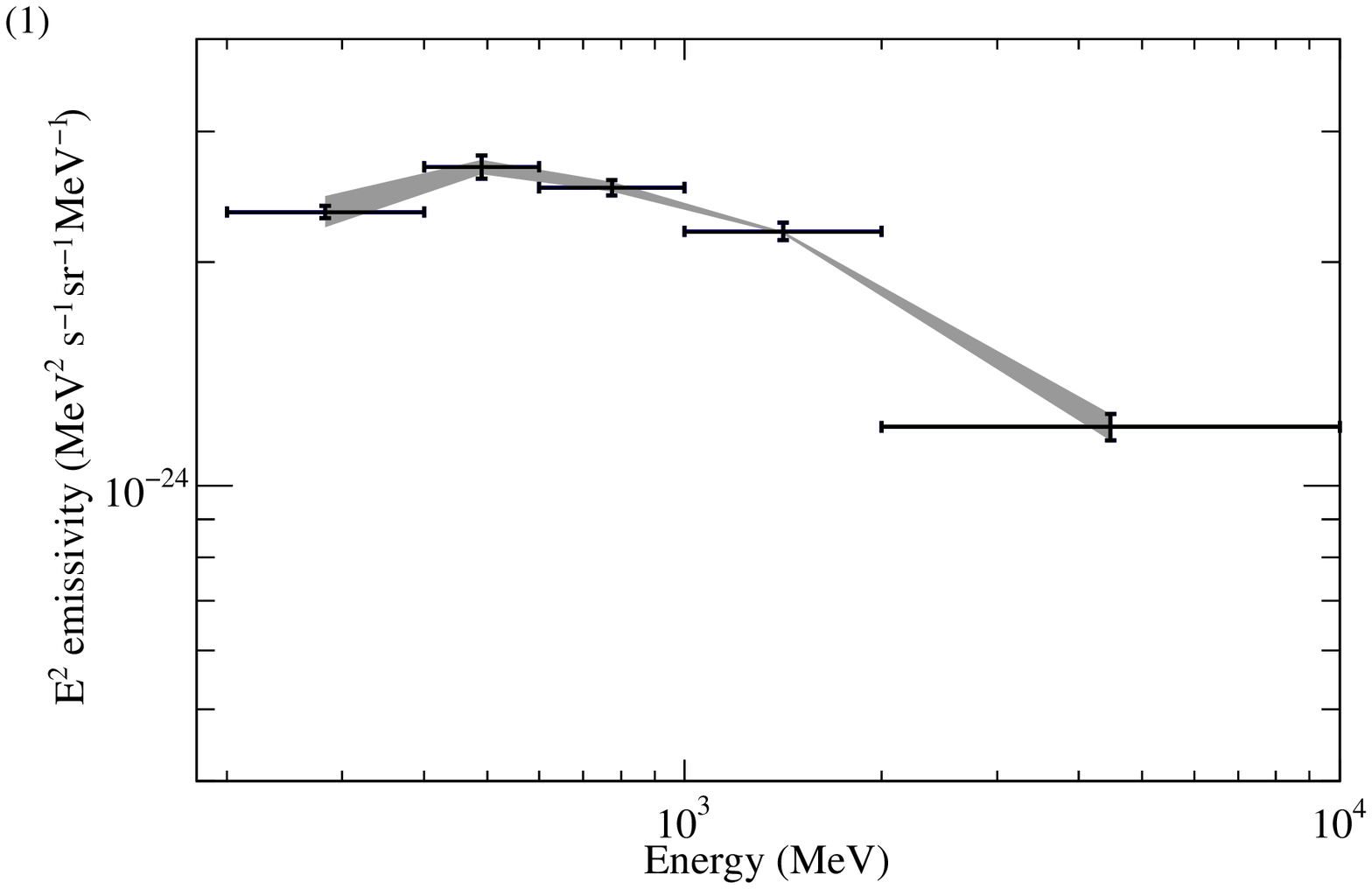}&\includegraphics[
width=0.48\textwidth]{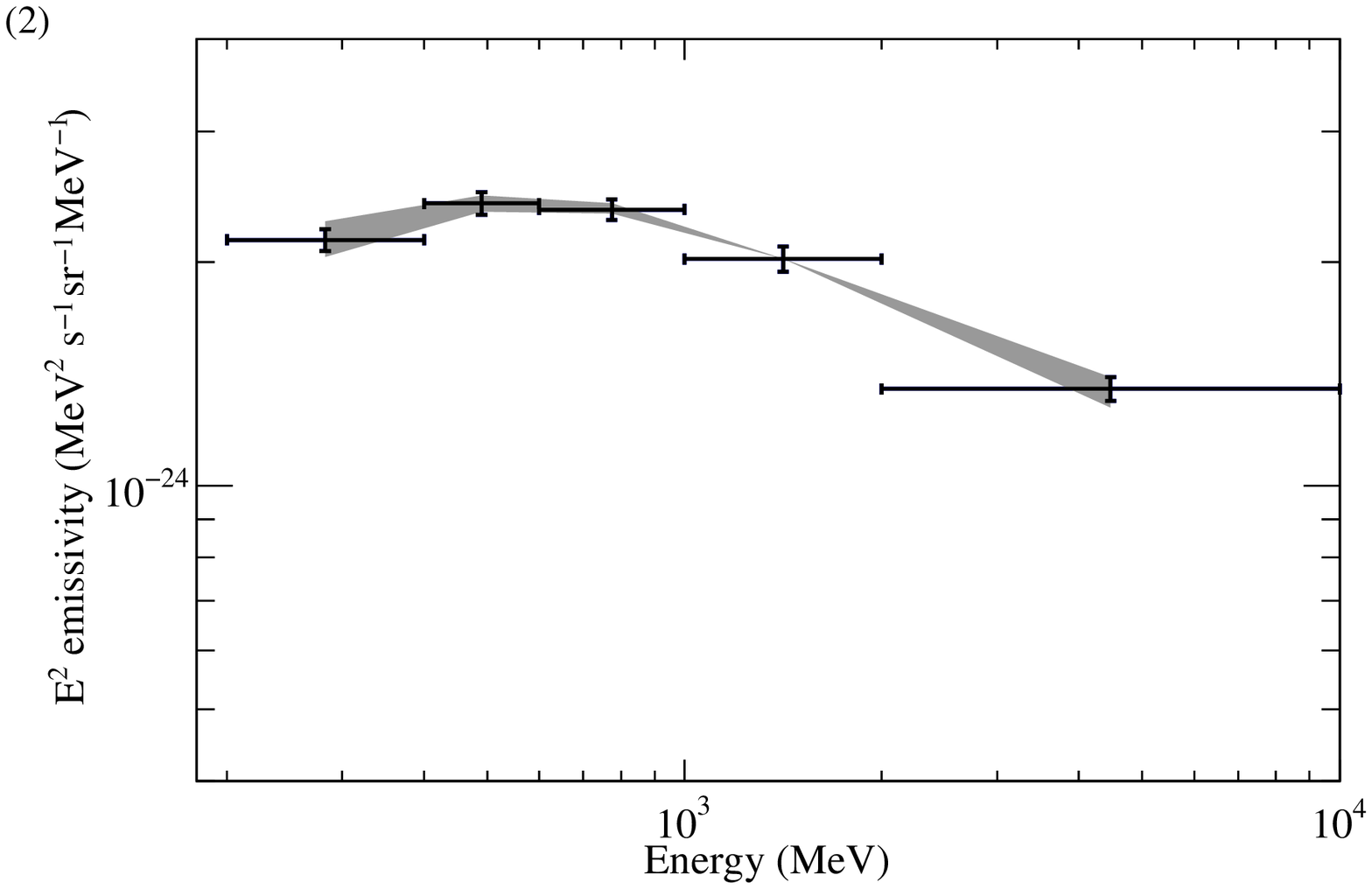}\\
\includegraphics[width=0.48\textwidth]{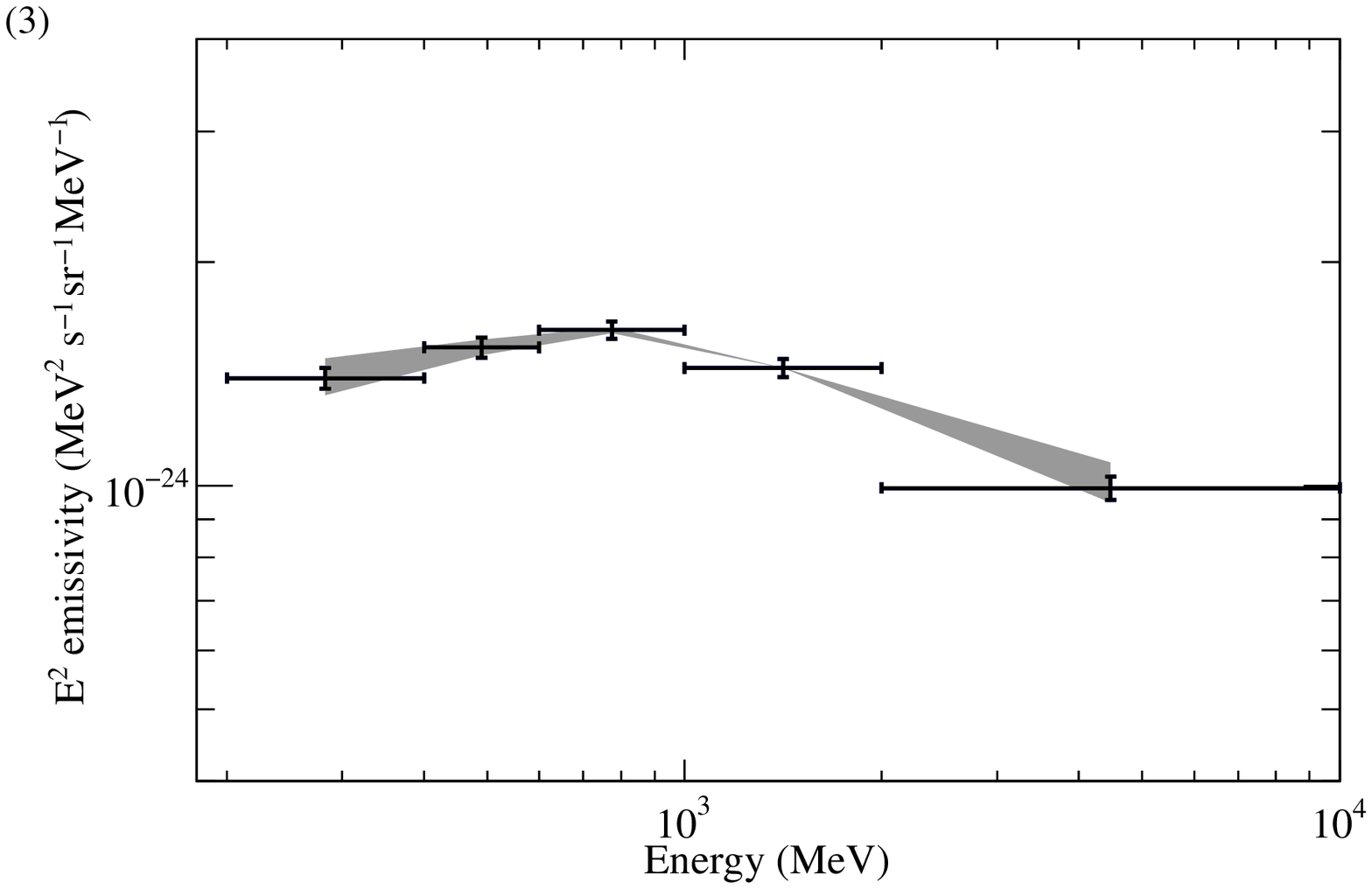}&
\end{tabular}
\caption{Emissivity spectra per $\hi$ atom as measured in the Gould Belt (1),
local arm (2), and Perseus arm (3) clouds. Horizontal bars mark the energy
bands, vertical bars show the statistical uncertainties on the measurement. The
shaded areas represent the systematic errors due to the uncertainties on the
event selection efficiency.}\label{aspec}
\end{figure*}
In Fig.~\ref{aspec} we report the emissivity spectra per $\hi$ atom measured in
the Gould Belt, the main part of the local arm and the Perseus arm. The
inclusion of the $\ebvres$ map in the fit does not have a strong impact on the
emissivities of the broadly distributed $\hi$ gas, which decrease by less than
10\% considering the interstellar reddening in analysis.

The results we obtained in the Gould-Belt and local-arm regions are consistent below 1 GeV with the measurement by \citet{egretcep}, obtained from EGRET observations of the region of Cepheus and Polaris. Above 1 GeV LAT measurements are $\sim40\%$ lower than those by EGRET. We will see below in \ref{locemiss} that LAT measurements are consistent with the \emph{a priori} expectations for the local $\hi$ emissivity: this result confirms that, as was already deduced from LAT observations of broader regions of the sky \citep{tapICRC,gevexc}, LAT measurements are not consistent with the GeV excess seen by EGRET, which was noticed also as an excess above 1 GeV in the emissivity of nearby $\hi$ complexes, as discussed in \citet{egretmono}.

Our spectra of the emissivity per $\hi$ atom are consistent with the results of an independent analysis carried out on LAT data to determine the local $\hi$ emissivity in a mid-latitude region of the third Galactic quadrant \citep{hiemiss}. The latter analysis investigated a different region of the sky, but encompassing $\hi$ complexes at $\lesssim 1$ kpc from the solar system, mostly located in a segment of the local arm. If we compare the present results in the main part of the local arm with those by \citet{hiemiss} we have excellent agreement. Therefore we have verified that CR proton densities smoothly vary on a few kpc scale around the solar system.

\subsubsection{Physical model}\label{physmodel}

We further compare our results with the predictions by GALPROP, a physical model of CR propagation in the Galaxy \citep[see e.g.][]{galprop98,galprop04,galprop07}. GALPROP solves the propagation equation for all CR species, given a CR source distribution and boundary conditions. Current GALPROP models assume a Galactocentric source distribution derived from that of pulsars \citep{strongXvar}. The distribution used by the model adopted for this work, called \texttt{54\_71Xvarh7S}, is given by Eq.~\ref{crsourceq}
\begin{equation}\label{crsourceq}
f(R) \propto \left(\frac{R}{R_\odot}\right)^\alpha \exp\left[-\beta \, \left(\frac{R-R_\odot}{R_\odot} \right)\right] 
\end{equation}
with $\alpha=1.25$, $\beta=3.56$ and $R_\odot=8.5$ kpc. A truncation is applied at $R=15$ kpc because we do not expect many CR sources in the outermost Galaxy. This choice of parameters results in a slightly flatter radial profile of CR densities than with the pulsar distribution.

The GALPROP model \texttt{54\_71Xvarh7S} is tuned to reproduce the \textit{in situ} measurements of CR spectra at the solar circle. The proton spectrum is derived from a compilation of direct measurements \citep{proton1,proton2,proton3}. The model includes the CR electron spectrum recently measured by the LAT \citep{eLAT}.

Once the propagation equation is solved, GALPROP computes the emissivity for stable secondaries, in particular $\gamma$ rays. The electron Bremmstrahlung component is evaluated using the formalism by \citet{ebremform} as explained in \citet{galpropbrem}. The emissivity due to $p$-$p$ interactions is evaluated using the inclusive cross sections as parametrized by \citet{kamaepp}. Following the method by \citet{dermera,dermerb}, the $p$-$p$ emissivity is increased to account for interactions involving CR $\alpha$ particles and interstellar He nuclei. This method provides an effective enhancement with respect to pure $p$-$p$ emissivities, often named the \textit{nuclear enhancement factor}, of $\nef \simeq1.45$. More recent calculations by \citet{mori09}, however, report values as large as $\nef \simeq1.75-2$ due to different CR spectral formulae \citep{honda04}, different ISM abundances and the inclusion of heavier nuclei both in CRs and in the ISM. Further theoretical developments are required to better constrain $\nef$, extending the predictions from $\gamma$ rays to other relevant messengers like antiprotons \citep{pamelaantip}.
 
\subsubsection{Emissivity in the Gould Belt}\label{locemiss}

\begin{figure}[!hbt]
\plotone{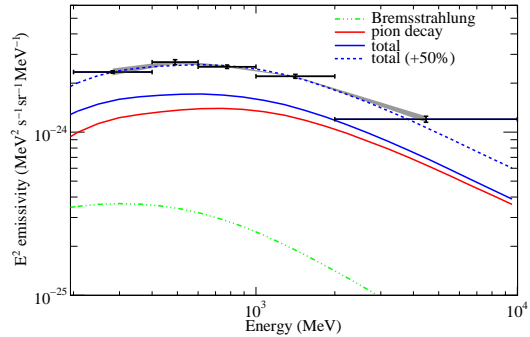}
\caption{$\hi$ emissivity spectrum in the Gould Belt, as shown in
Fig.~\ref{aspec}. The curves represent the predictions by GALPROP
\texttt{54\_71Xvarh7S}. The total emissivity from the model has been increased
by 50\% to reproduce our measurements.}\label{GBemiss}
\end{figure}

In Fig.~\ref{GBemiss}, we compare the emissivity spectrum per $\hi$ atom we
measured in the Gould Belt with the GALPROP predictions. We find the latter to
be $\simeq 50\%$ lower at all energies. As we have just discussed, a large part
of this excess ($\sim 30\%$) can be explained by the uncertainties in the
contribution from interactions involving CR and ISM nuclei other than protons.
The remaining $\sim 20\%$ excess can be explained by systematic uncertainties in
the CR proton spectra at the Earth ($\sim 20\%$), the $\nhi$ column-density
derivation, and the kinematical separation of emission from the outer Galaxy.

The $\hi$ emissivity in the Gould Belt clouds (within 300 pc from the solar system) is thus consistent with the hypothesis that the gas is interacting with CRs with the same spectra measured at Earth. Fig.~\ref{GBemiss} shows the GALPROP model scaled by $+50\%$ to highlight that the spectral shape is in good agreement with our results.

The $\hi$ spin temperature of 125 K (chosen to have a straightforward comparison with earlier analyses) is among the lowest values reported in the literature. A higher temperature would imply a higher emissivity, therefore a larger discrepancy with the GALPROP model (e.g. $\qhi{1}$ increases by another $5\%$ to $10\%$ if we take $T_S=250$ K as recently suggested by \citealt{dickey}).

\subsubsection{$\mathbf{\hi}$ emissivity gradient}

\begin{figure}[!hbt]
\plotone{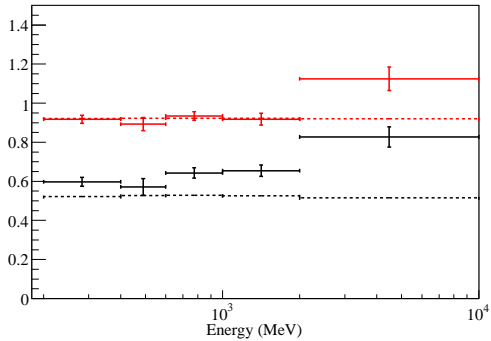}
\caption{Emissivities measured in the Perseus arm --black-- and in the local arm
--light (red)-- relative to those in the Gould Belt. Horizontal bars mark the
energy ranges, vertical bars show statistical uncertainties.  The dashed lines
represent the GALPROP predictions.}\label{emratiofig}
\end{figure}

It is evident from Fig.~\ref{aspec} that the $\hi$ emissivity decreases from the
Gould Belt to the Perseus arm, as expected from the declining distribution of
candidate CR sources in the outer Galaxy. Fig.~\ref{emratiofig}
shows the
emissivity ratios between the more distant regions and the Gould Belt.
Systematic errors due to the event selection efficiency are not relevant for
these ratios, because the emissivity spectra are similar.

The emissivity spectrum in the local arm is 10\% lower than in the Gould Belt. The GALPROP model predicts such a decrease because of the change in Galactocentric radius from the solar circle to the main part of local arm, located in this direction at $\sim 9.5$ kpc.

\begin{figure}[!htb]
\plotone{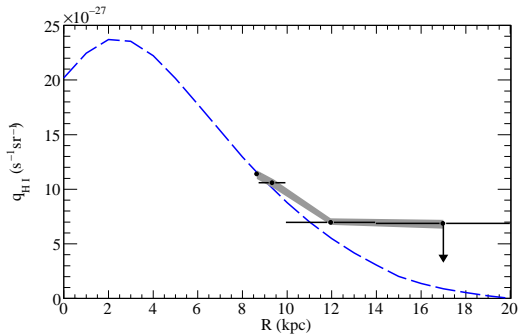}
\caption{Radial profile with Galactocentric radius of the $\hi$ emissivity
integrated between 200 MeV and 10 GeV. Black dots/horizontal bars mark the
ranges in kinematic distance encompassing the Gould Belt, the main part of the
local arm, the Perseus and outer arms (from left to right). Statistical
uncertainties on $\qhinull$ are smaller than the dot dimensions. The grey shaded
area shows the systematic uncertainties on the event selection efficiency. The
(blue) dashed line is the GALPROP prediction scaled up by 50\%.}\label{andyplot}
\end{figure}
A further decline in $\qhinull$ is expected between the local and
Perseus arms, but Fig.~\ref{emratiofig} shows that the measured $\qhinull$
gradient is significantly shallower than the GALPROP prediction.
In Fig.~\ref{andyplot} we
compare the $\hi$ emissivity integrated above 200 MeV predicted by GALPROP as a
function of Galactocentric radius with the values we measured in the four
regions defined for analysis, drawing the same conclusion. In this figure we
report the emissivity found in the outer arm, though considered only as an upper
limit because its determination is probably affected by faint sources (see
section \ref{anastep}).

The discrepancies between the measured and predicted gradients may be due to the large uncertainty in the CR source distribution. The SNR radial distribution across the Galaxy is very poorly determined because of the small sample available and large selection effects \citep{snrunc}. Distance and interstellar dispersion uncertainties also bias the pulsar distribution, in spite of the larger sample available \citep{lorimer04}. On the other hand, the CR diffusion parameters, derived from local isotopic abundances in CRs, may not apply to the whole Galaxy, as suggested by \citet{taillet}. Self absorption can also lead to a significant underestimate of $\nhi$ in the Perseus arm \citep{gibson}, thus to an overestimate of its $\gamma$-ray emissivity. Therefore, further investigation is needed to better understand the radial profile of the $\hi$ emissivity.

In Fig.~\ref{aspec}, the $\hi$ emissivity spectrum in the Perseus arm appears harder than expectations, thus suggesting that primary CR spectra vary across the Galaxy. We cannot, however, rule out energy-dependent systematic effects due to the separation power provided by the LAT PSF which strongly varies with energy, or a hardening due to contamination by hard unresolved point sources, like pulsars, clustering in the Perseus-arm structures.

\subsection{Cloud masses}

\subsubsection{CO}\label{codiscuss}

Because the $\gamma$-ray emission from molecular clouds is primarily due to $\hd$ and the molecular binding energy is negligible with respect to the energy-scale of the $\gamma$ radiation processes, the emissivity per $\hd$ molecule is twice the emissivity per $\hi$ atom. Under the hypothesis that the same CR flux penetrates the $\hi$ and CO phases of a cloud, we can assume that $\qco{\imath}=2 \, \xco{}_{,\,\imath} \cdot \qhi{\imath}$ in each region to derive the CO-to-$\hd$ conversion factor, $\xco$.

\begin{figure*}[!hbt]
\begin{tabular}{cc}
\includegraphics[width=0.48\textwidth]{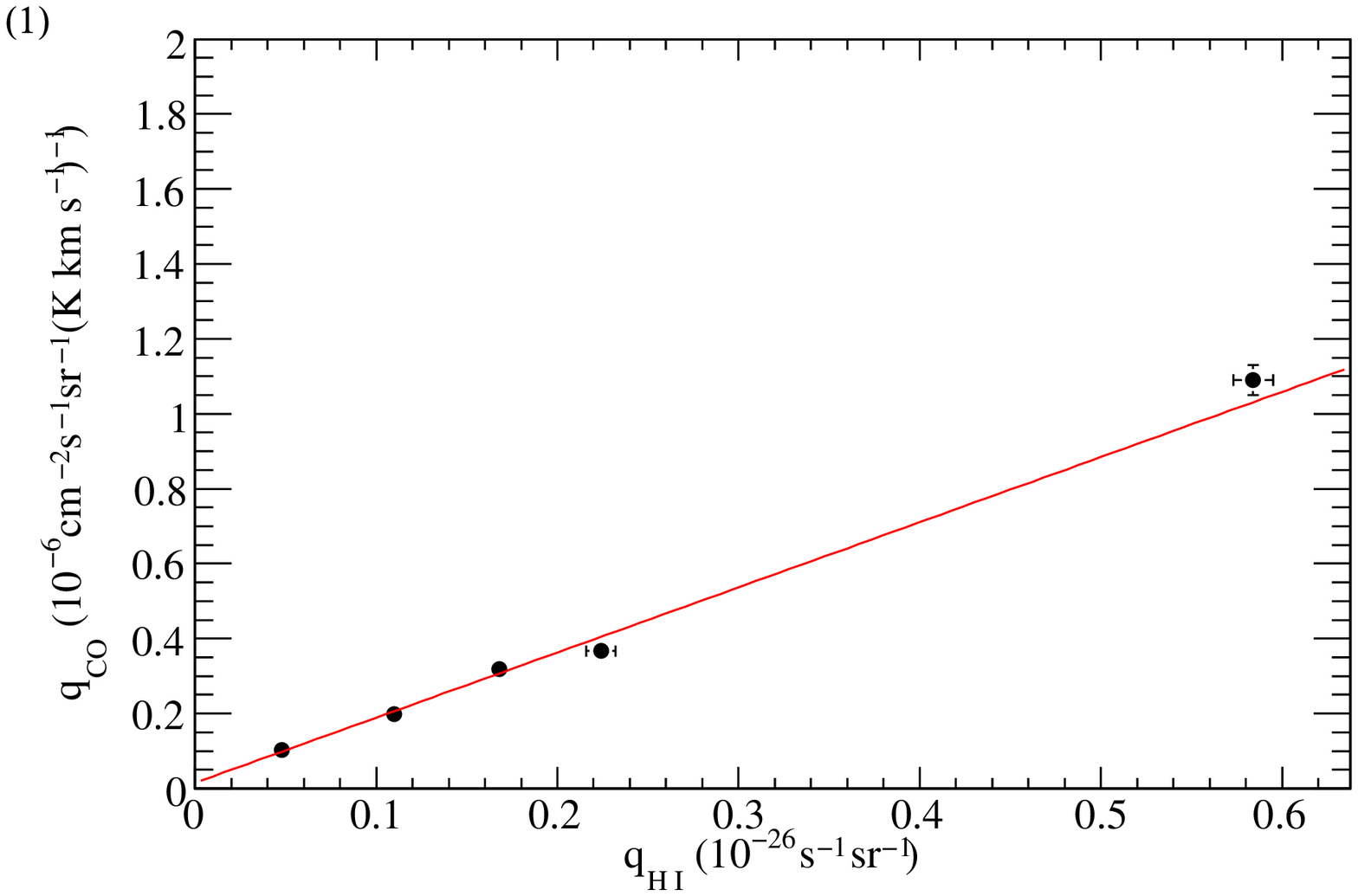}&\includegraphics[
width=0.48\textwidth]{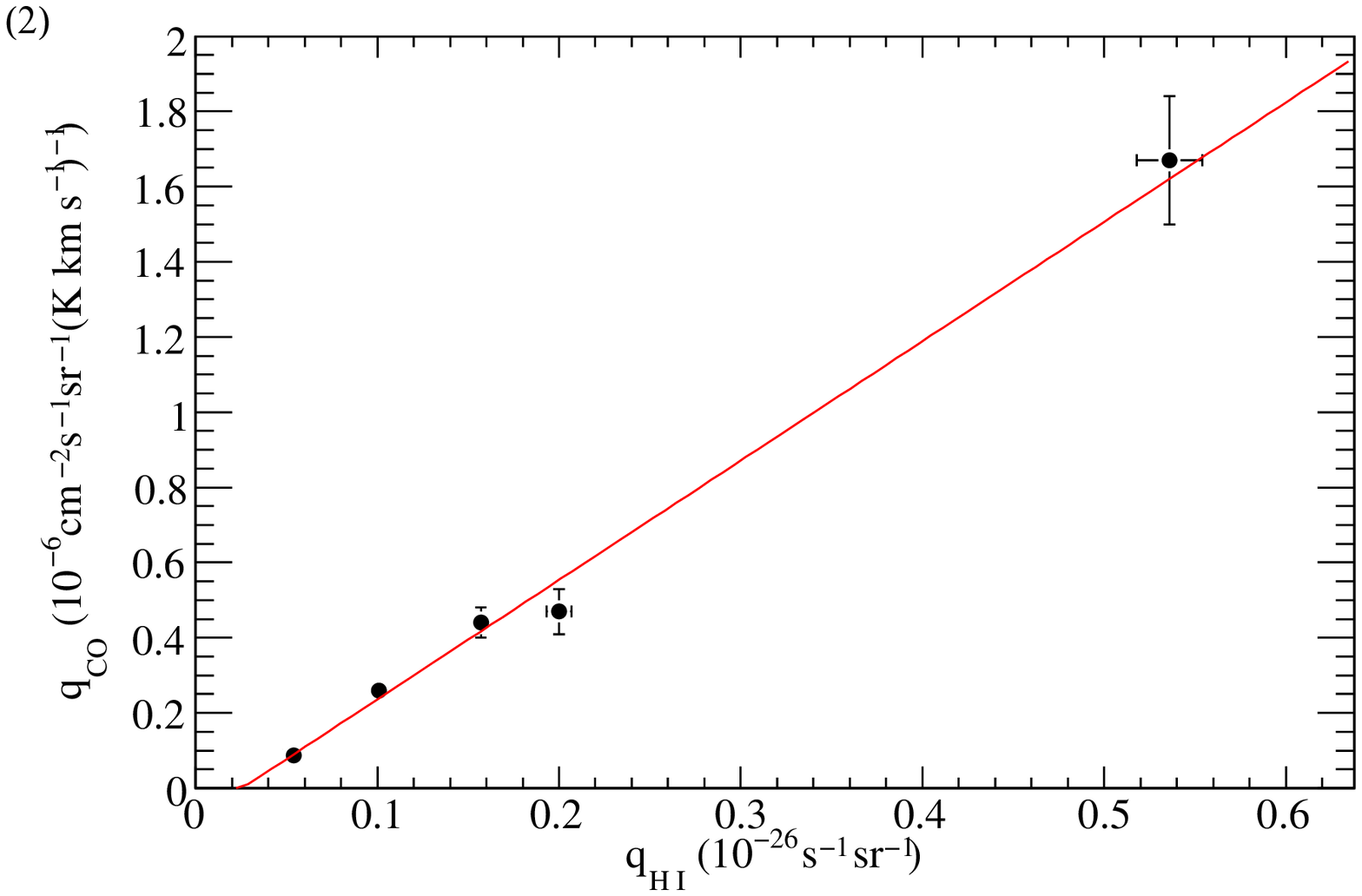}\\
\includegraphics[width=0.48\textwidth]{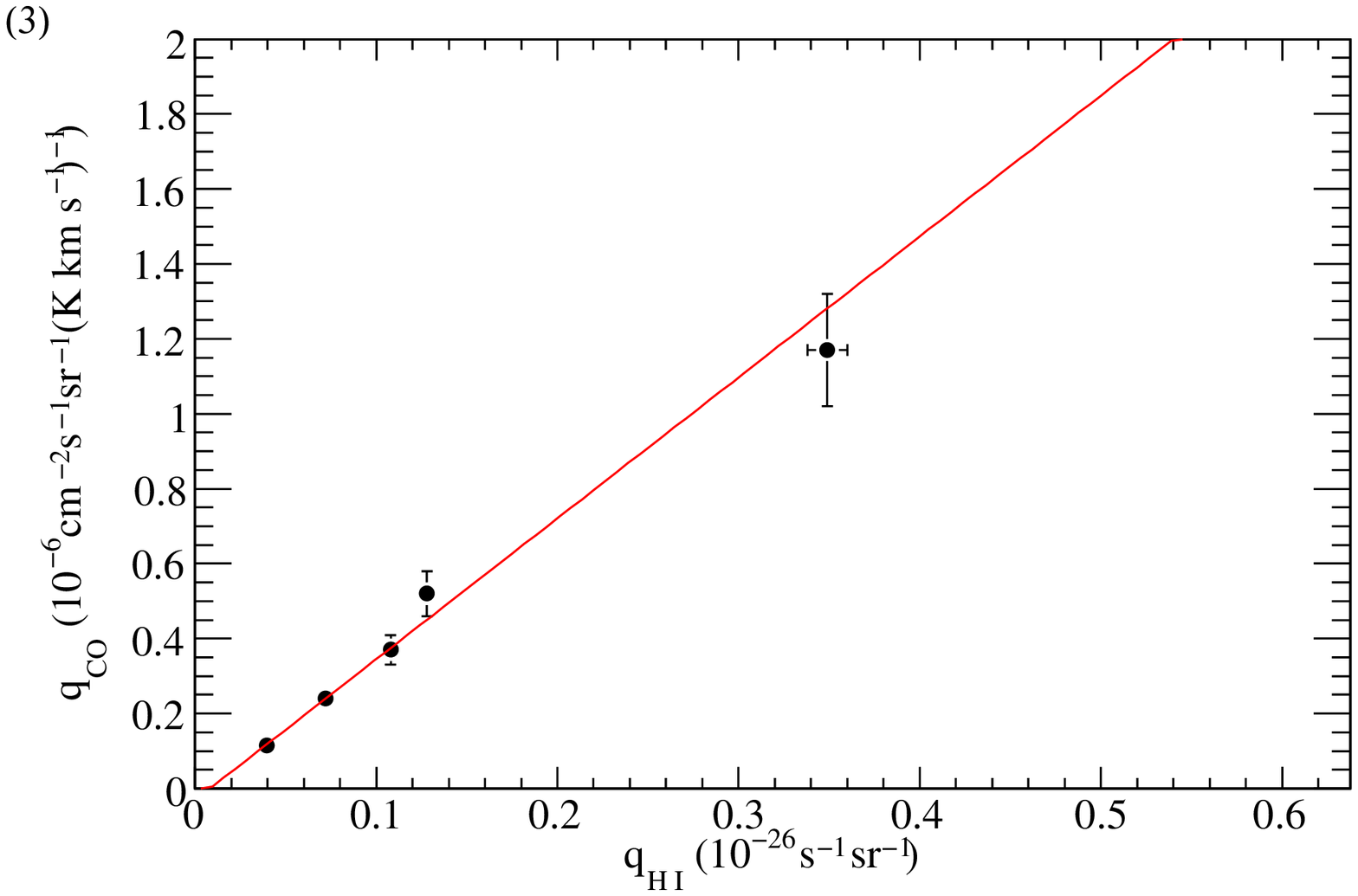}&
\end{tabular}
\caption{Correlation between the $\hi$ and CO emissivities obtained in the five
energy bands for each region under analysis: (1) Gould Belt, (2) local arm, (3)
Perseus arm. Error bars show the statistical uncertainties on $\qhi{\imath}$ and
$\qco{\imath}$. The (red) lines give the best linear fits.}\label{xrel}
\end{figure*}

We have performed a maximum likelihood linear fit  $\qco{\imath}=
\xco{}_{,\,\imath} \cdot 2 \qhi{\imath} + \overline{q}_\imath$ between the
$\qco{\imath}$ and $\qhi{\imath}$ values found in the various energy bands for
each region. We have taken into account the errors and covariances obtained from
the $\gamma$-ray fits for both $\qhinull$ and $\qconull$. Systematic errors due
to the event selection efficiency do not affect the derivation of the $\xco$
slope because the $\hi$ and CO emissivities have similar spectra. The results
are shown in Fig.~\ref{xrel},
and the best-fit parameters are reported in
Table~\ref{xtable}. We observe a good linear correlation between $\qhinull$ and
$\qconull$ that lends support to the assumption that CRs penetrate molecular
clouds uniformly to their cores \citep[still under debate, see e.g.][]{gabici}.
\begin{deluxetable}{c r @{$\pm$} l r @{$\pm$} l}
\tablewidth{0pt}
\tablecaption{Results of the linear fits between the $\hi$ and CO emissivities
in the different regions (1 -- Gould Belt, 2 -- local arm, 3 -- Perseus arm):
$\qco{\imath}= \xco{}_\imath \cdot 2 \qhi{\imath} +
\overline{q}_\imath$.\label{xtable}}
\tablecolumns{3}
\tablehead{
 & \multicolumn{2}{c}{$\xco{}$\tablenotemark{a}} &
\multicolumn{2}{c}{$\overline{q}$\tablenotemark{b}}
}
\startdata
1	& 0.87&0.05	& 0.015&0.012  \\
2	& 1.59&0.17	& -0.08&0.03 \\
3	& 1.9&0.2	& -0.03&0.03
\enddata
\tablenotetext{a}{Units: $10^{20}$ cm$^{-2}$ (K km s$^{-1}$)$^{-1}$).}
\tablenotetext{b}{Units: $10^{-6}$ cm$^{-2}$ s$^{-1}$ sr$^{-1}$ (K km
s$^{-1}$)$^{-1}$.}
\end{deluxetable}

\begin{figure}[!hbt]
\plotone{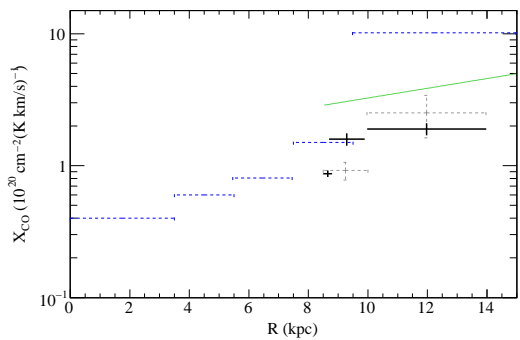}
\caption{ $\xco$ as a function of Galactocentric radius. The solid-line (black)
points represent our measurements: horizontal bars mark the ranges of kinematic
distance encompassing the Gould Belt and the local and Perseus arms (from left
to right), vertical bars show the statistical uncertainties on $\xco$ (errors
are statistical only, possible systematics are discussed in the text
in~\ref{codiscuss}). Dashed (blue) lines represent the values used in GALPROP by
\citet{strongXvar}. The solid (green) line shows the $\xco$ function determined
by \citet{Xjap} from CO data and virial masses (adapted to the rotation curve
assumed for our analysis). The dashed (grey) points show the previous EGRET
measurements in the region of Cepheus and Polaris \citep{egretcep}.}\label{XR}
\end{figure}
Fig.~\ref{XR} shows the $\xco$ variation with Galactocentric radius.
 Our measurements are consistent with previous $\gamma$-ray estimates in this
region of the sky \citep{egretcep}, but they are more precise, especially in the
outer Galaxy. For the segment of the Perseus arm near NGC 7538 we have lowered
the statistical uncertainty from $\sim 40\%$ to 10\%. The results suggest an
increase of $\xco$ in the outer Galaxy, as expected from the metallicity
gradient \citep[see e.g.][]{rolleston}. The $\xco$ measurements in external
galaxies indeed show a metallicity dependence possibly caused by CO
photodissociation and poor self-shielding in low-metallicity environments
\citep{israel1,israel2}.

Contamination from unresolved point sources with a spatial distribution closely related to that of the clouds is expected in star-forming regions which can produce young pulsars, supernova remnants, and massive binaries. This effect is unlikely in the Gould-Belt clouds (Cassiopeia, Cepheus, and Polaris), first because they form few high-mass stars, second because of the good linear resolution of the $\gamma$-ray maps of these nearby clouds. Their proximity ($\lesssim 300$ pc) and the $\sim 0.5^\circ$ angular resolution of the LAT in the higher energy band imply a linear resolution $\lesssim 3$ pc, which allows an efficient separation between diffuse emission and point sources. The contamination by point sources is limited for similar reasons in the nearby local arm, $\lesssim 1$ kpc away, but it cannot be clearly ruled out in the Perseus arm clouds which are known to form massive star clusters \citep[see e.g.][]{sandell}.

We cannot exclude separation problems between the $\gamma$-ray emission from the CO cores and their surrounding $\hi$ envelopes. The separation, based on the spatial distribution of the different phases, becomes less efficient with increasing distance due to the lower linear resolution. Moreover, we have verified that the presence of $\gamma$ rays associated with the dark-gas envelopes around the CO cores affects the determination of the CO-to-$\hd$ factor in more distant, not so well resolved, clouds (whereas the impact is negligible in the closer clouds). Excluding the $\ebvres$ map from the model yields a $\sim 30\%$ increase of $\xco$ in the Perseus arm. Unfortunately, the $\ebvres$ map is not reliable near the plane because of the confusion along the line of sight that prevents an adequate temperature correction and the removal of clusters of IR point sources, so this difference has to be considered as a systematic error on $\xco$ in the Perseus arm.

Whether the present $\xco$ gradient can be fully attributed to the metallicity
gradient, or partially to unresolved sources, $\hi$ and CO separation problems,
or gas not traced by $\hi$ and CO, needs further investigation, primarily at
higher resolution when more high-energy LAT data become available to profit from
the better angular resolution. For the moment, the fact that the present $\xco$
determination does not depend on energy (see Fig.~\ref{xrel}) suggests that
unresolved sources and separation of the different gas phases do not
significantly influence the result. The results shown in Fig.~\ref{XR} indicate
significantly smaller $\xco$ values in the outer Galaxy than those used by
\citet{strongXvar} in GALPROP and systematically smaller values than the
$\xco(R)$ relation determined by \citet{arimoto96} and \citet{Xjap} using CO
data and virial masses.

The $\xco$ values shown in Table~\ref{xtable} have been used to estimate cloud masses using Eq.~\ref{masseq}
\begin{equation}\label{masseq}
M = 2 \; \mu \; m_\mathrm{H} \; d^2 \; \xco \; \int \wco(l,b) \, \mathrm{d}\Omega
\end{equation}
where $d$ is the distance of the cloud, $m_\mathrm{H}$ is the H atom mass, and $\mu = 1.36$ is the mean atomic weight per H atom in the ISM. We did not use the kinematic distances inferred from CO surveys, but we adopted more precise estimates available in the literature. The results are given in Table~\ref{comasstable}. The errors include only the statistical uncertainties on $\xco$.

To investigate the discrepancies found between the different determinations of $\xco$ we calculated the virial masses for well-resolved clouds off the plane. The virial masses have been obtained from the CO velocity dispersion for a spherical mass distribution with density profile $\propto 1/r$, following Eq.~\ref{virmass}
\begin{equation}\label{virmass}
M=\frac{3}{2} \, \frac{r}{G} \, \sigma^2_v
\end{equation}
where $r$ is the cloud radius, $\sigma_v$ the velocity dispersion and $G$ is
Newton's constant. The velocity dispersion has been measured for each line of
sight and the average value in the sample has been taken as the characteristic
$\sigma_v$ in the cloud. This method limits the impact of the obvious velocity
gradients in these clouds. Because the virial mass heavily depends on the
estimate of the characteristic radius and on the cut-off applied in its
evaluation, we considered both the effective radius $r_A=\sqrt{A/\pi}$ (where
$A$ is the geometrical area of the cloud) and the intensity weighted radius
$\langle r \rangle=\left(\sum_\imath \wco{}_\imath \; r_\imath
\right)/\left(\sum_\imath \wco{}_\imath \right)$ (where $r_\imath$ is the
distance of pixel $\imath$ to the peak $\wco$ pixel). We truncated the
calculation at 1\% of the $\wco$ peak in both cases. We find that the virial
masses are systematically larger than the $\xco$ derived masses by a factor
1.5--3. This discrepancy in the nearby clouds is comparable to that shown in
Fig.~\ref{XR} between the $\gamma$-ray estimates of $\xco$ and the $\xco(R)$
function by \citet{Xjap} which relies on virial masses. The $\gamma$-ray
estimates are independent from the chemical, dynamical and thermodynamical state
of the clouds, but they can suffer from the limited resolution of $\gamma$-ray
surveys and the non-uniform penetration of CRs into the dense CO cores.
Conversely, the assumption of a spherical cloud in virial equilibrium against
turbulent motions is rather crude. Intrinsic velocity gradients and magnetic
pressure can easily bias the virial mass results.

\begin{deluxetable}{c c c r @{$.$} l r @{$\pm$} l r @{.} l r @{.} l r @{$\pm$}
l}
\tablewidth{0pt}
\tablecaption{Masses for specific clouds, complexes or regions obtained from CO
intensities and the $\xco$ values in Table~\ref{xtable}. For selected clouds we
also report virial masses and, in the Gould Belt, the dark-gas mass obtained
from the $\xebv$ conversion factor determined in~\ref{dgdiscuss}. All masses are
in units of $10^5 \, M_\odot$ and the errors include only the statistical
uncertainties on $\xco$ or $\xebv$.\label{comasstable}} 
\tablecolumns{7}
\tablehead{& $l$		&$b$		& \multicolumn{2}{c}{$d$ (kpc)}
& \multicolumn{2}{c}{$M_\mathrm{CO}$} & \multicolumn{2}{c}{$M_\mathrm{vir}
(r_A)$} & \multicolumn{2}{c}{$M_\mathrm{vir} (\langle r \rangle)$} &
\multicolumn{2}{c}{$M_\mathrm{dark}$} }
\startdata
Cepheus		& $[100,117]$	& $[6,22]$	& 0&3\tablenotemark{a}	
& 0.37&0.02	& 0&687	& 0&903	& 0.160&0.011\\
Polaris		& $[117,129]$	& $[18,30]$	& 0&25\tablenotemark{b}	
& 0.052&0.003	& 0&208	& 0&159	& 0.031&0.002\\
Cassiopeia	& $[117,145]$	& $[2,18]$	& 0&3\tablenotemark{a}	
& 0.61&0.03	& 0&893	& 1&062	& 0.34&0.02\\
Gould Belt 	& $[100,145]$	& $[-15,30]$	& 0&3			
& 1.47&0.08	\\
NGC 7538	& $[107,115]$	& $[-5,5]$	& 2&65\tablenotemark{c}	
& 20&2		\\
NGC 281		& $[120,125]$	& $[-9,-5]$	& 3&0\tablenotemark{d}	
& 0.79&0.08	& 1&205	& 1&047\\
Perseus arm 	&$[100,145]$	& $[-10,10]$	& 3&0			
& 57&6		
\enddata
\tablenotetext{a}{\citet{greniercepcas}}
\tablenotetext{b}{\citet{polaris}}
\tablenotetext{c}{\citet{moscadelli}}
\tablenotetext{d}{\citet{sato}}
\end{deluxetable}

\subsubsection{Dark gas}\label{dgdiscuss}

In order to quantify the significance of the correlation between the
$\gamma$-ray intensities and the $\ebvres$ map, we have repeated the last step
of \ref{anastep} without including it in the analysis. The corresponding test
statistics, $\mathrm{TS}=2 \Delta (\ln \like)$, obtained in the five energy
bands are given in Table~\ref{dgTS}. With the addition of two free parameters
($\qebv$ and a spectral index), in the null hypothesis that there is no
$\gamma$-ray emission associated with the $\ebvres$ map TS should follow a
$\chi^2$ distribution with two degrees of freedom. Therefore, the correlation
between $\gamma$ rays and $\ebv$ residuals is verified at a confidence level
$>99.9\%$ in all energy bands.
\begin{deluxetable}{l c}
 \tablewidth{0pt}
\tablecaption{$\mathrm{TS}=2 \Delta (\ln \like)$ for the inclusion of the
$\ebvres$ map in the fit in the different energy bands. \label{dgTS}}
\tablecolumns{2}
\tablehead{
 \colhead{energy range (GeV)} & \colhead{TS}}
\startdata
0.2--0.4	& 53.8 \\
0.4--0.6	& 124 \\
0.6--1		& 74.6 \\
1--2		& 91.8 \\
2--10		& 38.2 \\
\enddata
\end{deluxetable}

The magnitudes of the dust masses and dust IR emission are too low to explain this correlation by CR interactions with dust grains or their thermal radiation. However, the correlation can be explained by CR interactions in normal gas that is not accounted for in the $\nhi$ and $\wco$ maps.

In section \ref{codiscuss} we have used the $\gamma$-ray emissivities per $\hi$ atom and $\wco$ unit to calibrate the CO-to-$\hd$ conversion factor, following a well-established method. We can use a similar procedure to correlate the $\gamma$-ray emissivities per $\hi$ atom and per $\ebvres$ unit in the well resolved Gould-Belt clouds (see Fig.~\ref{dgmap}) where the spatial association between the $\hi$, CO and $\ebvres$ maps allows locating the dark gas in the absence of kinematical information.

\begin{figure}[!hbt]
\plotone{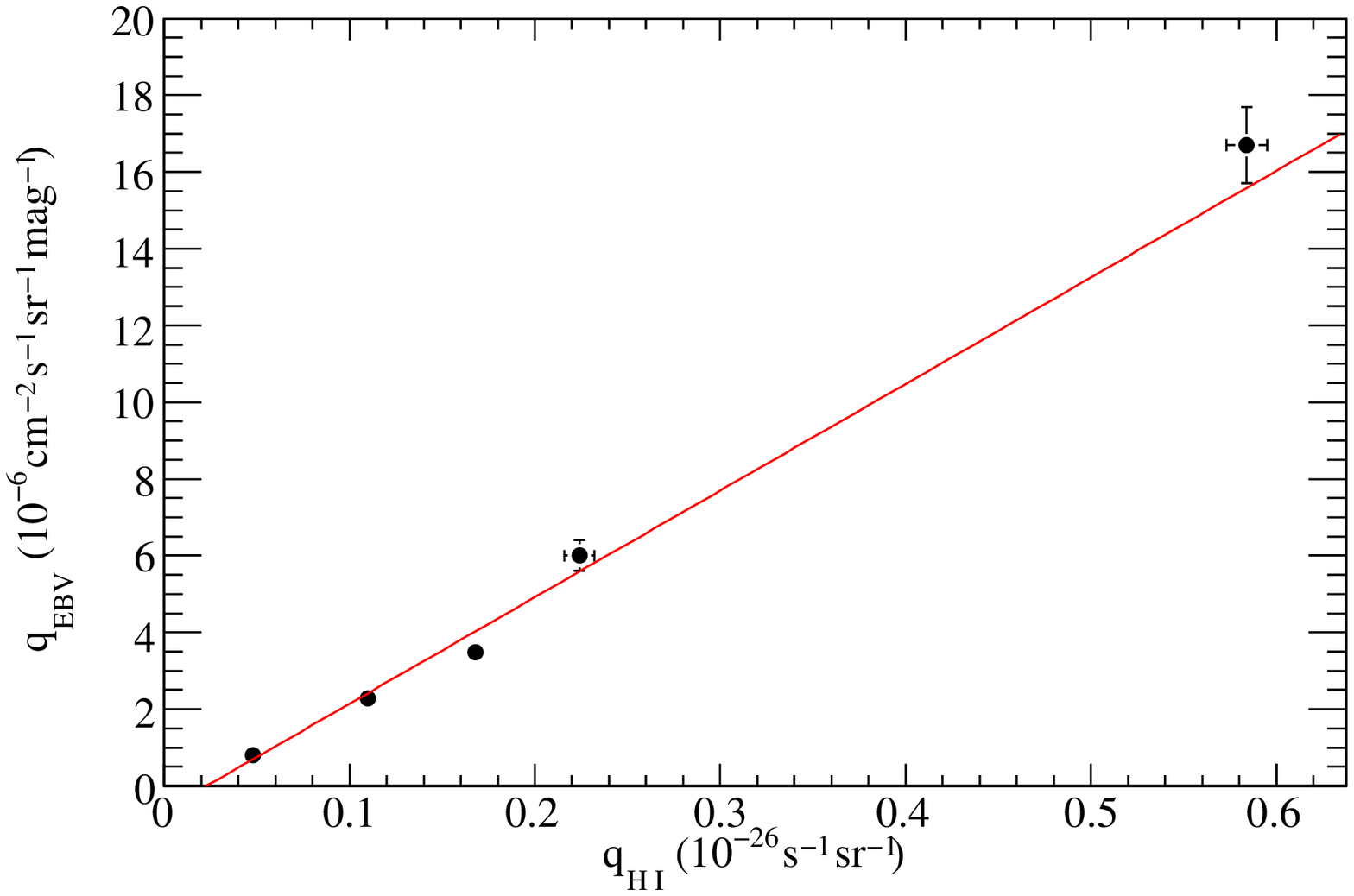}
\caption{Correlation in the Gould Belt between the $\hi$ emissivities and the
emissivities per unit of $\ebvres$. Error bars show the statistical errors
obtained on both emissivities in the five energy bands. The (red) line shows the
best linear fit.}\label{Xebv}
\end{figure}
The $\qhi{1}$ and $\qebv$ emissivities found in the five energy
bands exhibit a tight correlation (Fig.~\ref{Xebv}). As we did for CO, we fitted
a linear relation, $\qebv=\xebv \cdot \qhi{1} +\overline{q}$, using a maximum
likelihood method taking into account the errors and covariances of the
emissivities. The results are $\xebv=(28 \pm 2) \times 10^{20}$ cm$^{-2}$
mag$^{-1}$ and $\overline{q}=(-0.6 \pm 0.2) \times 10^{-6}$ cm$^{-2}$ s$^{-1}$
sr$^{-1}$ mag$^{-1}$. The good linear correlation implies similar spectra for
the $\gamma$-ray emission from gas seen in the $\hi$ emission line and that
associated with the excess reddening, thus confirming the need for normal
additional gas to explain the correlated excess of both $\gamma$ rays and dust
at the interface between the $\hi$ and CO emitting phases of the Gould Belt
clouds.

Using the $\xebv$ factor in these clouds, we can calculate the additional gas mass and compare it to the molecular mass seen in CO. We restrict the comparison to the CO mass, because the more diffuse $\hi$ clouds are difficult to separate from the background $\hi$ disc extending to intermediate latitudes. To estimate the dark mass, we use only the positive residuals in the $\ebvres$ map. As discussed in section \ref{dgsec}, the small negative residuals associated with the CO cores are likely related with local variations in the dust temperature or dust-to-gas ratio. The results are given in Table~\ref{comasstable}. The errors include only the statistical uncertainties on $\xebv$. The additional mass in the Gould-Belt clouds appears to be 40\% to 60\% of the CO-bright mass. We note that the sum of the dark and CO mass is closer to the virial one. We also note that FIRAS and SIMBA dust spectra in the Cepheus flare led to an independent estimate of its total mass, $M=(0.43\pm0.18)\times 10^5 M_\odot$ \citep{mmclouds}, which relates well with the total (CO plus dark) mass $M=(0.53\pm0.02)\times 10^5 M_\odot$ we have obtained in $\gamma$ rays.

\section{Summary}

We have analysed the interstellar $\gamma$-ray emission observed by the \emph{Fermi} LAT in the region of Cassiopeia and Cepheus, successfully modeling the $\gamma$-ray data as a linear combination of contributions arising from different gas complexes towards the outer Galaxy.

The separation has allowed us to verify that the $\gamma$-ray emissivity of local atomic gas is consistent with production by interactions with CRs with the same spectra as those measured near the Earth, but confirms the higher pion-decay contribution relative to
some of the estimates in the literature, as found in \citet{hiemiss}. This can be plausibly attributed to uncertainties in the local CR spectra, either in the measurement or from differences between the direct measurements and local interstellar space.

Thanks to the correlation between an excess of dust and of $\gamma$-ray emission, with a spectrum equivalent to that found for the atomic and molecular gas, we have verified the presence of an excess of gas not properly traced by the standard $\nhi$ and $\wco$ maps. In the nearby Gould-Belt clouds, the dark gas forms a layer between the $\hi$ and CO phases and it represents about $50\%$ of the mass traced in the CO-bright molecular cores.

The CR-density gradient in the outer Galaxy appears to be flatter than expectations based on the assumption that CRs are accelerated by SNRs as traced by pulsars. It is also possible that the CR spectrum in the Perseus arm is harder than in the local arm. This hardening, which needs confirmation at high resolution with more LAT data to limit the potential contamination by hard unresolved point sources, could be linked to CR diffusion not far from their sources.

We have measured $\xco$ in several regions from the Gould Belt to the Perseus arm. The $\gamma$-ray estimates are independent of the chemical and thermodynamical state of the gas and also from assumptions on the virial equilibrium of the clouds. They correspond to a significant but moderate increase of $\xco$ with Galactocentric radius outside the solar circle, from $(0.87 \pm 0.05) \times 10^{20}$ cm$^{-2}$ (K km s$^{-1}$)$^{-1}$ in the Gould Belt to $(1.9\pm 0.2) \times 10^{20}$ cm$^{-2}$ (K km s$^{-1}$)$^{-1}$ in the Perseus arm.

\acknowledgments
$\phantom{a}$\\
The \textit{Fermi} LAT Collaboration acknowledges generous ongoing support
from a number of agencies and institutes that have supported both the
development and the operation of the LAT as well as scientific data analysis.
These include the National Aeronautics and Space Administration and the 
Department of Energy in the United States, the Commissariat \`a l'Energie Atomique
and the Centre National de la Recherche Scientifique / Institut National de Physique
Nucl\'eaire et de Physique des Particules in France, the Agenzia Spaziale Italiana
and the Istituto Nazionale di Fisica Nucleare in Italy, the Ministry of Education,
Culture, Sports, Science and Technology (MEXT), High Energy Accelerator Research
Organization (KEK) and Japan Aerospace Exploration Agency (JAXA) in Japan, and
the K.~A.~Wallenberg Foundation, the Swedish Research Council and the
Swedish National Space Board in Sweden.

Additional support for science analysis during the operations phase is gratefully
acknowledged from the Istituto Nazionale di Astrofisica in Italy and the
Centre National d'\'Etudes Spatiales in France.

We thank T.~H.~Dame for providing moment-masked CO data including from some
observations not yet published.

\end{document}